\documentclass{aa}  

\usepackage{subcaption}
\usepackage{float}
\usepackage{xcolor}
\usepackage[hyphenbreaks]{breakurl}
\usepackage{natbib,twoopt}
\usepackage[breaklinks]{hyperref} 
    
\bibpunct{(}{)}{;}{a}{}{,}             
\definecolor{cobalt}{rgb}{0.06, 0.2, 0.65}
\hypersetup{
  colorlinks,
  citecolor=cobalt,
  linkcolor=[rgb]{0.8, 0.2, 1.0},
  urlcolor=cobalt,
}
\makeatletter
  \newcommandtwoopt{\citeads}[3][][]{\href{http://adsabs.harvard.edu/abs/#3}%
    {\def\hyper@linkstart##1##2{}%
     \let\hyper@linkend\@empty\citealp[#1][#2]{#3}}}
  \newcommandtwoopt{\citepads}[3][][]{\href{http://adsabs.harvard.edu/abs/#3}%
    {\def\hyper@linkstart##1##2{}%
     \let\hyper@linkend\@empty\citep[#1][#2]{#3}}}
  \newcommandtwoopt{\citealtads}[3][][]{\href{http://adsabs.harvard.edu/abs/#3}%
    {\def\hyper@linkstart##1##2{}%
     \let\hyper@linkend\@empty\citealt[#1][#2]{#3}}}
  \newcommandtwoopt{\citeyearads}[3][][]%
    {\href{http://adsabs.harvard.edu/abs/#3}
    {\def\hyper@linkstart##1##2{}%
     \let\hyper@linkend\@empty\citeyear[#1][#2]{#3}}}
\makeatother

\usepackage{graphicx}
\usepackage[varg]{txfonts}

\usepackage[normalem]{ulem}

\begin{document} 

   \title{Structure and evolution of cosmic voids within DESI DR1}

   \author{Óscar Monllor-Berbegal\inst{1}\thanks{oscar.monllor@uv.es},
                Susana Planelles\inst{1,2},
                Vicent Quilis\inst{1,2}
          }

   \institute{Departament d’Astronomia i Astrofísica, Universitat de València,
              46100 Burjassot (València), Spain
         \and
            Observatori Astron\`omic, Universitat de Val\`encia, E-46980 Paterna (València), Spain}

   \date{Received X XX, XXXX; accepted X XX, XXXX}

  \abstract
  {Ongoing Stage IV galaxy surveys such as DESI or Euclid offer an unprecedented amount of information to study the large-scale structure of the Universe by mapping tens of millions of galaxies. Within the context of cosmic voids, existing and future catalogues will improve our understanding of the void structure and evolution as well as the cosmological model constraints derived from their statistics.}
  {With the recent DESI Data Release 1, our aim is to obtain a new void catalogue that allows us to study the properties of cosmic voids at different redshifts, hence providing insight into not only their structure but also their evolution.}
  {We selected the DESI BGS and LRG samples for this study. Before the void-finding process, the data went through a preprocessing pipeline tackling redshift-space distortions and reconstructing the mildly non-linear velocity field. After that, the new \texttt{AVISM} void finder was applied to the preprocessed data, obtaining two void catalogues in different redshift bins ($0.1<z<0.3$  and $0.4<z<0.8$), both within a $1400^3 \, \mathrm{cMpc}^3$ cube.}
  {We successfully recovered \mbox{$N_\text{BGS} = 1114$} and \mbox{$N_\text{LRG} = 216$} clean voids with sizes up to $\sim70 \, \mathrm{cMpc}$. The void size function, 3D shape variables, density profiles, and linear bias were obtained and analysed. Both the resulting size function and the density profiles are in agreement with previous theoretical studies. The expected mass depletion in void evolution is shown by the evolution of density profiles. A constant linear bias inside $r<R_e$ was obtained, with mean values of $1.31 \pm 0.02$ and $2.14 \pm 0.04$ for BGS and LRG voids, respectively. Both void catalogues are publicly released.}
   {}

   \keywords{Cosmology: observations -- Galaxies: general -- large-scale structure of Universe -- Methods: numerical -- Methods: data analysis}

   \titlerunning{Structure and evolution of cosmic voids within DESI DR1}
   \authorrunning{Monllor-Berbegal, Planelles, Quilis}

  \maketitle
%

\section{Introduction}
\label{s:intro}

The cosmic web is formed by dense interconnected walls, filaments, and clusters \citep{bond1996filaments,cautun2014evolution}, and also by vast underdense regions that we know as cosmic voids. 
The shape, size distribution, and inner structure of voids contain a considerable amount of information on the underlying cosmology that can be exploited to constrain cosmological parameters \citep{lavaux2012precision,pisani2015counting,pisani2019cosmic,paz2023guess,contarini2024perspective, song2024void, fernandez2025constraining}. In particular, due to their low mass content they are dark energy dominated from an early stage in cosmic time and, hence, sensitive to its properties \citep{bos2012darkness, sahlen2019cluster}. For the same reason, voids provide an optimal ground for modified theories of gravity to test their hypotheses and neutrino masses to be constrained \citep{cai2015testing, baker2018void, davies2019cosmological, perico2019cosmic, bayer2021detecting, contarini2021cosmic}.

Regarding galaxy formation and evolution, voids constitute pristine environments where galaxies display different properties from those found in filaments and clusters \citep[e.g.][]{ricciardelli2014star, ricciardelli2017morphological, rosas2022revealing, curtis2024properties, argudo2024morphologies, perez2025galaxy}. In this direction, a more complete perspective of how galaxies form and evolve can be attained by studying how they behave according to these different environments. 

Achieving a large enough and accurate observational void catalogue is key to studying these scenarios. In this regard, only in the past decades have surveys such as the 2dF Galaxy Redshift Survey  \citep[2dF;][]{colless20012df} or the Sloan Digital Sky Survey \citep[SDSS;][]{abazajian2009seventh} allowed for the acquisition of large enough void samples \citep{pan2012cosmic, sutter2012public, sutter2014voids}. The advent of more advanced Stage IV surveys, such as the ongoing Dark Energy Spectroscopic Instrument \citep[DESI;][]{aghamousa2016desi} and Euclid \citep{euclid2025overview}, will cover the largest volumes so far with similar or better tracer densities, thereby providing an unprecedented level of information. The opportunity to obtain better constraints on the cosmological model and the study of the large-scale structure (LSS) is now open. In particular, such large volumes open the possibility of increasing void statistics considerably. A first step in this direction was carried out in \citet{rincon2025desivast}, where the first public catalogue of voids within DESI was presented and released.

In this work, our aim is to further contribute to the exploitation of this new era of galaxy surveys in order to disentangle the nature of cosmic voids, both in their structure and evolution. The recent DESI Data Release 1 \citep[DR1;][]{desi2026DR1} data were preprocessed and prepared for the void-finding process carried out by the \texttt{AVISM} code \citep{monllor2025avism}. The preprocessing pipeline corrects redshift-space distortions and reconstructs the velocity field in order to allow the void finder to work with its full capabilities. Voids were identified within two $1400^3 \; \mathrm{cMpc}^3$ cubes, one from the bright galaxy \citep{hahn2023desi} and the other from the luminous red galaxy \citep{zhou2023target} samples. The resulting two void catalogues were studied after obtaining the void size functions, histograms of their 3D structure relevant quantities, density profiles, and linear bias. Furthermore, for the sake of reproducibility and also to facilitate studies that could benefit from using these catalogues (e.g. on galaxy evolution), we publicly release the data.

The rest of the manuscript is structured as follows. In Sect.~\ref{s:survey} we provide details about the survey and the galaxy sample preparation. In Sect.~\ref{s:methodology}, the preprocessing pipeline and the void-finding details are explained. The main results, namely the void statistics and profiles, are presented and discussed in Sect.~\ref{s:results}. We summarise and conclude the work in Sect.~\ref{s:conclusions} and publicly release the catalogues in Sect.~\ref{s:data_availability}. Appendix~\ref{app:RSD_AP_impact} quantifies the impact of redshift-space distortions and the Alcock--Paczynski effect on void properties, Appendix~\ref{app:vel_recon_test} describes and tests our velocity reconstruction methodology, and Appendix~\ref{app:obs_mock_equivalence} justifies our approach to estimating the linear bias in DESI voids.

\section{Survey}
\label{s:survey}

In this section, we describe both the DESI survey and the DR1 data (Sect.~\ref{s:desi_dr1}), the sample details and selection (Sect.~\ref{s:sample_selection}), and the choice of fiducial cosmology that allowed us to define the final data cubes (Sect.~\ref{s:fid_cosm}).

\subsection{DESI DR1}
\label{s:desi_dr1}

DESI is a state-of-the-art, five-year spectroscopic redshift survey designed to map the expansion history of the Universe and explore the fundamental nature of dark energy. At its final stage, it will provide a $14,000 \, \mathrm{deg}^2$ sky coverage within $z \lesssim 4$, revealing the redshift of over 40 million objects. The survey targets distinct tracer populations over different cosmic epochs: the bright galaxy survey \citep[BGS;][]{hahn2023desi} probes the low-redshift Universe ($z < 0.4$) and can be understood as a
deeper version of the galaxy target selection for the SDSS main galaxy sample \citep{strauss2002spectroscopic}; luminous red galaxies \citep[LRGs;][]{zhou2023target} span intermediate redshifts ($0.4 < z < 1.1$); emission-line galaxies \citep[ELGs;][]{raichoor2023target}, which are the largest tracer sample, go further beyond ($0.6 < z < 1.6$); and quasars \citep[QSOs;][]{chaussidon2023target} extend up to $z \approx 4$, making them the sparsest sample \citep{aghamousa2016desi}. The DESI main survey consists of 9929 dark tiles, 5676 bright tiles, and 2657 backup tiles \citep{schlafly2023survey}, thus yielding three main programmes to be completed during the five years. The survey was validated in \citet{desi2024validation}. 

Currently, the most complete public data release is DR1 \citep{desi2026DR1}, which represents a meaningful
fraction of the full DESI survey: the dark survey is 29.0\% complete, the bright survey is 41.3\% complete, and the backup programme is 5.2\% complete, representing all data acquired during the first 13 months. A public void catalogue has already been published using DESI DR1 BGS data \citep{rincon2025desivast}, though the authors limited their work to $z \le 0.24$ and do not provide a comprehensive statistical analysis of their void population. 

\subsection{LSS catalogues and final galaxy samples}
\label{s:sample_selection}

In the DR1 public release, the DESI collaboration also provides the LSS catalogues \citep{ross2025construction}, which are designed to carry out clustering analyses. To achieve this, only proper spectroscopic observations are considered, and data are restricted to the quantities necessary to calculate clustering statistics. To fully describe the 3D mask of the survey, the observed data are paired with ‘randoms’ that Poisson-sample the observed space. To match the galaxy and random catalogues, weights are applied to both. Sub-sampling is also applied to the randoms in order to account for effects such as the survey footprint, completeness, instrument performance, and systematic errors.  

In this work, we restricted our void-finding analysis to the DESI DR1 LSS BGS and LRG samples. Since they lie in different redshift (cosmic time) bins, our results allowed us to study the evolution of cosmic voids. We did not go beyond into the ELG sample because it lies at a higher redshift ($z \sim 1.1$), where the lack of completeness combines with the fact that voids are smaller and less empty at this redshift, making their detection significantly harder. In fact, as can be seen in the visual impression of Sect.~\ref{s:results}, the void-finding process for the LRG sample was already complex. QSOs, on the other hand, can be directly ruled out because of their poor number density.

For the BGS case, a lower redshift cut of $z_\text{min} = 0.1$ was considered in order to minimise selection bias \citep{hahn2023desi,desi2024III}, obtain a relatively constant number density, and avoid the volume close to the observer where Poisson variance dominates. The upper limit cut ($z_\text{max} = 0.3$) was chosen to physically separate BGS galaxies from LRGs and as a compromise between having a large enough volume to obtain a significant void catalogue and enough resolution to carry out the void-finding process with precision. The LRG redshift range was chosen to have a similar comoving distance coverage to the BGS and also to capture the range in which its number density is almost constant. Thus, the following redshift ranges were used: \mbox{$z_\text{BGS} \in (0.1,0.3)$, $\; z_\text{LRG} \in (0.4,0.8)$}. 

\subsection{Fiducial cosmology and data cubes}
\label{s:fid_cosm}

Assuming a fiducial cosmology is mandatory in order to convert redshift to distances, carry out the correction of redshift-space distortions, reconstruct the velocity field, or calculate the theoretical fit for the void size function. Thus, we chose the following values as our fiducial cosmology: \mbox{$\Omega_{\mathrm{m},0}= 0.31$}, \mbox{$\Omega_{\Lambda,0} = 0.69$}, \mbox{$H_0 = 67.8 \, \mathrm{km \, s^{-1} \mathrm{Mpc}^{-1}}$}, and $\sigma_8 = 0.81$, in agreement with \citet{collaboration2020planck} results and compatible with DESI DR1 baryon acoustic oscillations (BAO) constraints \citep{adame2025desi}, both within the $\Lambda$CDM model.

The void finder (see Sect.~\ref{s:voidfinder}) relies on the use of a regular grid with cubic cells as the resolution elements to discretise the quantities that feed the void-finding process. The grid dimensions can be adjusted depending on the physical application. For survey data with complex volume coverage, a mask is used to label cells in the computational grid that are not within the survey coverage. How accurate is the description of the survey geometry depends directly on numerical resolution, with the perfect choice being a compromise between computational expenses and a good approximation to the complex survey geometry. 
\footnote{Note that there are no restrictions on the morphology of identified voids, besides the resolution limit due to the cell size.}
 
For the sake of simplicity, we chose to study a cubic volume discretised in a regular grid. 
For both the BGS and LRG samples, this cube must fulfil the following criteria: i) maximise the fraction of galaxies from the sample inside it; ii) minimise the number of cells marked outside of the survey volume. This approach allowed us to save computational resources while providing a large-enough volume to carry out our study.

Using our fiducial cosmology, we calculated the 3D Cartesian position of all BGS galaxies and LRGs belonging to the previously described samples. Following the above criteria, the following cube boundaries delimit the final input volumes provided to the void finder (in units of cMpc):

\begin{equation}
\begin{split}
\label{eq:BGS_cube}      
      -1200 < \; & X_\text{BGS}  < 200, \\
      -700 < \; & Y_\text{BGS}  < 700, \\
       -200 < \; &Z_\text{BGS}  < 1200,  
\end{split}
\end{equation}
for the BGS and 
\begin{equation}
\begin{split}
\label{eq:LRG_cube}
      -2800 < \; & X_\text{LRG}  < -1400, \\
      -700 < \; & Y_\text{LRG}  < 700, \\
      -700 < \; & Z_\text{LRG}  < 700,  
\end{split}
\end{equation}
for the LRGs. Note that such volume selection was performed after carrying out the redshift-space distortion corrections and velocity reconstructions described in the next section. 

Finally, although we provided the void finder with an input volume of $1400^3 \, \mathrm{cMpc}^3$ in both cases the effective volume --- that is, the observed volume (defined by those regions populated by the randoms described in the previous section) --- reduces to $\approx 38\%$ and $\approx 50\%$ of the total cubic volume for the BGS and LRGs, respectively.

\section{Methodology}
\label{s:methodology}

This section describes the main parts of the preprocessing pipeline designed to prepare the DESI data to be handled by the \texttt{AVISM} void finder \citep{monllor2025avism}. First, we calculate the (clustering) linear bias for our galaxy samples (Sect.~\ref{s:gal_bias}). Then, we describe our redshift-space distortion (RSD) correction (Sect.~\ref{s:RSDs}) and velocity reconstruction (Sect.~\ref{s:reconstruction}) methods. Finally, a description of the particularities of the void-finding approach is provided (Sect.~\ref{s:voidfinder}).

\subsection{Galaxy bias}
\label{s:gal_bias}

It is well known that dark matter haloes and galaxies are more clustered than the overall matter content of the Universe \citep[e.g.][]{tinker2010large}. This evidence leads to the concept of bias ($b$), which in its simplest formulation establishes a proportionality between the galaxy distribution and the total mass distribution:
\begin{equation}
    \label{eq:bias}
    \delta_g \equiv n_g / \overline{n}_g - 1 = b (\rho / \overline{\rho}-1) \equiv b \delta ,
\end{equation}
where $\delta_g$ is the galaxy density contrast, $n_g$ is the galaxy number density, $\delta$ is the matter density contrast, and $\rho$ is the matter density. In both cases, the line above the density (e.g. $\overline{n}_g$) denotes the average over the whole Universe volume\footnote{In cosmological simulations and surveys, the average would be carried out over the available volume.}. This relation is accurate enough only on the largest (linear regime) scales \citep{desjacques2018large}. 

The fact that galaxies are biased must be taken into account when studying voids identified using them as matter tracers, as their sizes, shapes, or profiles may vary with respect to what is identified with the whole matter content \citep[e.g.][]{pollina2017linearity, contarini2019cosmological}. Moreover, a proper value of $b$ is necessary to accurately correct RSDs and reconstruct the underlying velocity field.

The value of the linear bias for BGS and LRGs in DESI depends on the redshift range and magnitude cuts applied to the samples. Since we used a particular redshift range and did not apply any cut of this kind, our sample biases may significantly differ from the standard fiducial values adopted by the DESI collaboration \citep[e.g.][]{adame2025desi} to carry out the BAO analysis. To overcome this issue, one can use clustering arguments to calculate the effective value of the linear bias. For this purpose, we used the \texttt{pycorr}\footnote{\url{https://github.com/cosmodesi/pycorr}} package \citep{pyrecon2, pyrecon1} to calculate the monopole ($\xi_0$) of the two-point correlation function (2PCF) for both galaxy samples in redshift space. Combined with the 2PCF associated with the matter linear power spectrum ($\xi_\mathrm{lin}$)\footnote{Our fiducial values for the cosmological parameters were used to calculate $\xi_\mathrm{lin}$. Note that $\sigma_8$ is perfectly degenerate with $b$.} calculated at each sample's effective redshift, one can estimate the linear bias value by fitting \citep{kaiser1987clustering}
\begin{equation}
    \label{eq:bias_calc_2PCF}
    \xi_0(d) = \bigg( b^2 + \frac{2 b f}{3} + \frac{f^2}{5}\bigg) \,\xi_\mathrm{lin}(d),
\end{equation}
where $b$ is the linear bias clustering value, $f \approx \Omega_\mathrm{m}(z)^{0.6}$ the linear growth rate \citep{peebles2020principles}, and $d$ the separation in redshift space. We obtained both $\xi_0$ and $\xi_\mathrm{lin}$ in $d \le 100 \,\mathrm{cMpc}/h$, evaluated Eq.~\eqref{eq:bias_calc_2PCF}, and averaged the values of $b$ over $d \in (20, 60) \,\mathrm{cMpc}/h$, where it displays an almost constant behaviour. The correlation function covariance matrix was obtained by jackknife estimation, providing the statistical error.

The following values were obtained: \mbox{$b_\mathrm{BGS} = 1.22 \pm 0.06$} and \mbox{$b_\mathrm{LRG} = 2.03 \pm 0.11$} for the BGS and LRGs, respectively. In Sect.~\ref{s:bias_in_voids} we also estimate the values of the linear bias inside BGS and LRG voids and compare them to the previously described clustering values.

\subsection{Redshift and geometrical distortions}
\label{s:RSDs}

Before applying any void finder to an observational survey, linear \citep{kaiser1987clustering, hamilton1998linear} and non-linear \citep{jackson1972critique} RSDs and the Alcock--Paczynski \citep[AP;][]{alcock1979evolution} effect must be taken into account, as they impact void centring, shapes, and, ultimately, void distribution \citep[e.g.][]{hamaus2015probing, cai2016redshift, nadathur2019beyond, hamaus2020precision, correa2021redshift, correa2022redshift, radinovic2024alcock}. While linear RSDs, known as the Kaiser effect, arise due to the coherent motion (infall) of galaxies on large scales ($\gtrsim15 \, \mathrm{Mpc}/h$) and result in a squashing along the line of sight, non-linear distortions, known as fingers of God (FoG), happen on smaller scales ($\lesssim5 \, \mathrm{Mpc}/h$) and are produced by the high velocity dispersion of galaxies inside clusters, causing an elongation along the line of sight. On the other hand, the AP effect appears when we convert the redshift and angular position of each galaxy to its comoving 3D position. If the true cosmology differs from the fiducial one assumed to carry out the transformation, a geometrical distortion appears on the galaxy map. 

For this work, we tackled RSDs but did not correct distortions due to the AP effect. Mild differences between the fiducial and true cosmology produce negligible distortions compared to RSDs \citep{ballinger1996measuring}. In Appendix~\ref{app:RSD_AP_impact} we describe a test in which we obtained that the impact of the AP effect on void properties was subdominant when compared to RSDs, for a moderate AP perturbation. Therefore, the impact of the AP effect was not considered to deliver a significant enough perturbation to be taken into account in our data preparation and analysis. Below, we detail the different procedures carried out to overcome the linear and non-linear RSDs.

\paragraph{Fingers of God:} Non-linear RSDs can intrude on voids and contaminate the structure with mass that, in reality, lies outside of the void boundary. Several approaches have been presented in the literature that attempt to amend the small-scale RSDs. The most common methodology for BAO reconstruction \citep{eisenstein2007improving, padmanabhan20122, ried2024velocity, chen2026extensive} and void-galaxy cross-correlation \citep{hamaus2016constraints, nadathur2019beyond, hamaus2020precision, radinovic2023euclid, radinovic2024alcock} analysis involves erasing the signal from FoG by smoothing the data on sufficiently large scales. For applications where higher spatial resolution is needed ($\sim 5 \, \mathrm{Mpc}/h$), a more careful treatment may be required. For instance, \cite{carrick2015cosmological}, \cite{tully2015galaxy}, and \cite{lilow2021constrained} find galaxy groups with the friends-of-friends \citep[FoF;][]{huchra1982groups} algorithm, placing galaxies at the mean redshift of their associated group and effectively suppressing the elongation along the line of sight. A similar methodology is presented by \citet{kuchner2021cosmic}, the authors of which propose to identify groups of galaxies and collapse them radially until all galaxies fall inside a $2\times R_{200}$ distance from the cluster centre.

Inspired by the aforementioned studies, and following a similar approach to \citet{tegmark2002power,tegmark2004three}, we implemented a FoG correction by using a FoF algorithm that identifies groups of galaxies and corrects the elongation in galaxy positions along the line of sight. Instead of collapsing all galaxies to the mean position or relying on the calculation of $R_{200}$, we rectified the FoG by ensuring that the standard deviation of positions along the line of sight ($\sigma_{\parallel}$) in all groups is the same as the one in the transverse direction ($\sigma_{\perp}$). In this regard, for each of the identified groups, we defined the following parameter:
\begin{equation}
    \label{eq:shape_param}
    \vartheta = \min\Big(C_1 \frac{\sigma_{\perp}}{\sigma_{\parallel}}, 1\Big)\,, 
\end{equation}
which quantifies the degree of sphericity and whose maximum value is 1, as we only wanted to correct deviations from sphericity along the line of sight. To avoid a FoG overcorrection, we introduced the constant $C_1 > 1$. We tested that $C_1 = 4/3$ yielded proper results.

Given $\vartheta$ for a given group, the corrected position ($\mathbf{s}'$) of each galaxy belonging to that group is
\begin{equation}
    \label{eq:FoG_correction}
    \mathbf{s}' = \mathbf{s} - (1-\,\vartheta)\, \Delta\mathbf{s}_{\parallel}\, ,
\end{equation}
where $\mathbf{s}$ is the comoving redshift-space position and \mbox{$\Delta\mathbf{s}_{\parallel} = (\Delta\mathbf{s}\cdot\mathbf{\hat{r}})\mathbf{\hat{r}}$}\,, with $\mathbf{\hat{r}}$ being the unit vector along the line of sight and $\Delta \mathbf{s}$ the galaxy position relative to the group centre.

Regarding the identification of groups and the application of Eq.~\eqref{eq:FoG_correction}, one first has to define the minimum number of members (galaxies) to constitute a group and then properly choose the linking length to recover them. For the first, we set $N_\text{min} = 10$ as the minimum number of galaxies to form a group. For the second, one has to take into account the fact that galaxy groups are smeared along the line of sight according to\footnote{The first-order Doppler effect is a good approximation only if \mbox{$v_r/c \ll 1$}.}
\begin{equation}
    \label{eq:smearing}
    \mathbf{s} \approx \mathbf{x} + \frac{v_r}{aH}\mathbf{\hat{r}}\, ,
\end{equation}
where $\mathbf{x}$ is the comoving real-space position of a galaxy, $v_r$ is the velocity along the line of sight, $H=H(t)$ is the Hubble parameter, and $a$ the scale factor. By Eq.~\eqref{eq:smearing} it becomes clear that, in order to capture the elongated groups, the linking length along the line of sight must be larger than the one for the transverse direction: $l_{\parallel} = f_{e}\,l_{\perp}$ with $f_e > 1$. We can estimate a typical elongation scale to physically motivate $f_e$. For a sufficiently relaxed cluster of galaxies of a given virial mass and radius, one can obtain a typical circular velocity of galaxies moving within it as
\begin{equation}
    \label{eq:circular_vel}
    v_c = \sqrt{\frac{G M_\text{vir}}{R_\text{vir}}}\, ,
\end{equation}
which, for a typical mass of $M_\text{vir} = 5\times{10^{14}\,M_\odot}$ and radius of $R_\text{vir} = 2\,\mathrm{Mpc}$ yields $v_c \sim 1000\, \mathrm{km \, s^{-1}}$. With this value of the velocity, we can estimate the maximum elongation of the FoG effect on this cluster by assuming that, in the worst case, the velocity lies completely along the line of sight and, thus, $\frac{v_r}{aH} \approx 15 \, \mathrm{cMpc}$\footnote{Here, we assumed $z \approx 0$.}, yielding $f_e \approx 7$. In practice, such a high $f_e$ value is not necessary for two reasons: i) such a large elongation only happens in the most massive cases; ii) precisely in these massive clusters there will be sufficient galaxy counts to connect them and form the FoF group, even for lower linking-length values.

For this work, we chose $f_e = 3$ as a compromise between being able to capture the elongated groups and avoiding percolation (bridging). We finally chose the linking lengths such that the transverse one is equal to the mean distance of a galaxy to its closest neighbour, yielding $l_\perp \approx 1.5 \, \mathrm{cMpc}$ and $l_{\parallel} \approx 5 \, \mathrm{cMpc}$ for the BGS sample and $l_\perp \approx 6 \, \mathrm{cMpc}$ and $l_{\parallel} \approx 18 \, \mathrm{cMpc}$ for the LRGs. Two particles (galaxies in this case) are joined together and considered `friends' if $d_{\parallel} < l_{\parallel}$ and $d_{\perp} < l_{\perp}$ \citep[similar to][]{reid2010cosmological}, with $d_{\perp} = \sqrt{d^2 - d^2_{\parallel}}$, where $d$ is the pair separation and $d_{\parallel}$ ($d_{\perp}$) its component along the line-of-sight (transverse) direction.

\paragraph{Kaiser effect:} In order to deal with linear RSDs we followed the standard BAO reconstruction approach. In particular, we used the \texttt{pyrecon}\footnote{\url{https://github.com/cosmodesi/pyrecon}} package that implements the \citet{burden2015reconstruction} algorithm and solves the Zel’dovich equation in redshift space:
\begin{equation}
    \label{eq.zeldovich_recon}
    \nabla \cdot \mathbf{\Psi} + \frac{f}{b} \nabla \cdot[(\mathbf{\Psi}\cdot \mathbf{\hat{r}})\,\mathbf{\hat{r}}] = - \frac{\delta_g}{b}\, ,
\end{equation}
 where $\mathbf{\Psi}$ is the Lagrangian displacement field (see Appendix~\ref{app:vel_recon_test}) and $\delta_g$ the galaxy density contrast in redshift space. \citet{chen2026extensive} provides a thorough description of the details involved in the reconstruction process and evaluates its accuracy for the different samples from the DESI DR1 data. Taking this into account, Eq.~\eqref{eq.zeldovich_recon} was solved by applying the \citet{burden2015reconstruction} algorithm on a $512^3$ grid for both the BGS and LRG samples. The appropriate value for the smoothing scale of this method was taken from \citet{desi2024III}.

Once the displacement field was recovered, galaxy positions were shifted by
\begin{equation}
    \label{eq:kaiser_term}
    \mathbf{\Psi}_\text{RSD} = - f (\mathbf{\Psi}\cdot \mathbf{\hat{r}})\,\mathbf{\hat{r}}\, ,
\end{equation}
which is the linear RSD contribution. We thus obtained the real-space configuration.

Finally, as a test to assess the effectiveness of the RSD treatment (especially with regard to the FoG contribution), we calculated the two-dimensional 2PCF, $\xi(d_{\parallel}, d_{\perp})$ before and after performing the correction steps. We used the standard estimator by \citet{landy1993bias}:
\begin{equation}
    \label{eq:2pcf}
    \xi(d_{\parallel}, d_{\perp}) = 1 + \bigg(\frac{N_R}{N_D}\bigg)^2 \frac{DD}{RR} - 2 \frac{N_R}{N_D}\frac{DR}{RR}\, , 
\end{equation}
where $N_D$ ($N_R$) is the number of data (random) points, and $DD$, $DR$, and $RR$ are the data--data, data--random, and random--random pair counts binned in the $d_{\parallel} - d_{\perp}$ plane. We calculated the 2PCF within a maximum pair distance of $15 \, \mathrm{cMpc}$ in both the parallel and perpendicular directions.

\begin{figure}[h!]
\centering 
\includegraphics[width=1\linewidth]{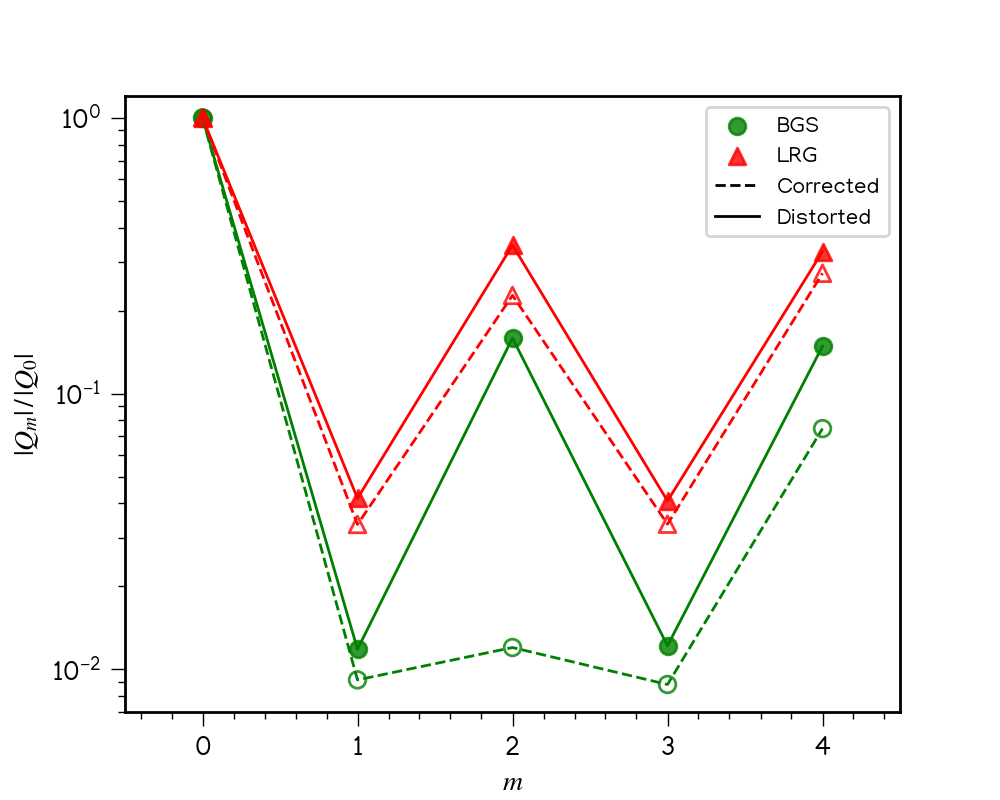}
\caption{Multipole expansion of the corrected (dashed) and distorted (solid) 2PCF for the BGS (green circles) and LRG (red triangles) samples. The aperture radius is $\Delta R = 15 \, \mathrm{cMpc}$.}
\label{fig:2pcf}
\end{figure}

After that, we obtained a multipole expansion of $\xi$ according to the methodology presented in \citet{schneider1997aperture}:
\begin{equation}
    Q_m (\Delta R) = \int_{\Delta R} R \,\mathrm{d}R \int_0^{2\pi} \mathrm{d}\phi \, e^{im\phi} \xi(R, \phi),
\label{eq:multipole}
\end{equation}
where $Q_m$ are the different multipole moments obtained within a $\Delta R$ aperture and $\xi(R, \phi)$ is the 2PCF in polar coordinates. The aperture was fixed to $\Delta R = 15\, \mathrm{cMpc}$ according to the $d_{\parallel} - d_{\perp}$ binning. If the FoG correction works properly, the anisotropies of the 2PCF, that is, the multipole moments beyond the monopole ($Q_{m}$ for $m>0$), should considerably decrease their magnitude. The results are presented in Fig.~\ref{fig:2pcf}. In the BGS case, the quadrupole ($m=2$) is vastly reduced, as is the $m=4$ moment to a lesser extent. On the other hand, for the LRG sample, even after applying the RSD corrections the anisotropies in the 2PCF remain significant; this is mainly due to the ineffectiveness of the FoF detection because of the poor sample density, which only allows the identification and correction of the largest groups. Accordingly, we expect a worse quality (more noise-dominated) void catalogue within the LRG data. 

\subsection{Velocity reconstruction}
\label{s:reconstruction}

With real-space positions recovered after tackling linear and non-linear RSDs, we could readily apply most void finders available in the literature, as they do not require the velocity field to carry out their void-finding process \citep[see][for example, for a list of publicly available void finders]{contarini2026era}. However, for the purpose of this work, we employed the \texttt{AVISM} code  \citep{monllor2025avism} to generate the void catalogues from the galaxy samples. This void finder uses not only density but also dynamical information through the divergence of the velocity field. In the linear regime, these two quantities are directly related \citep{peebles2020principles}:
\begin{equation}
    \label{eq:linear}
    \nabla_{\mathbf{x}} \cdot \mathbf{v} = - f a H\delta(\mathbf{x},t)\, , 
\end{equation}
where $\mathbf{x}$ is the real-space comoving position, and $a$ is the scale factor. Due to this relation, the void finder can accept an input without the velocity, as it can obtain an estimation of the velocity field via \mbox{Eq.~\eqref{eq:linear}}. Nevertheless, due to the proportionality, the velocity divergence extracted from that equation does not carry more information beyond the density field.

Several methodologies exist to go beyond linear theory. For instance, Bayesian frameworks \citep[e.g.][]{jasche2019physical, kitaura2021cosmic} recover non-linear densities and velocity fields by forward modelling the initial linear density fluctuations until the evolved density field matches the data. Likewise, machine-learning techniques \citep[e.g.][]{wu2021cosmic,ganeshaiah2023large,lilow2024neural} trained with $N$-body simulations are recovering increasingly better (non-linear) reconstructions of the physical quantities. In \citet{monllor2025avism}, we generate an all-sky void catalogue for the local Universe by applying \texttt{AVISM} to non-linear data reconstructed with the first methodology. 

Such reconstructions, nevertheless, are still not available (to our knowledge) for DESI data, mainly due to the complexity (large volume, limited sky coverage, variable completeness, etc.) of the survey. This is the reason why we opted for the quasi-linear path-interchange Zel’dovich approximation method \citep[PIZA;][]{croft1997reconstruction} to reconstruct the velocity field (see e.g. \citealt{elyiv2015cosmic} and \citealt{sartori2026back}, for similar applications). In Appendix~\ref{app:vel_recon_test}, we provide a brief summary of the method and outline the main particularities of our implementation. A test against galaxies from a cosmological simulation from which halo peculiar velocities are already known is also presented. Reconstructed and real velocities are well correlated; they lie on the 1:1 ideal relation with a Pearson correlation coefficient of $\approx 0.74$. 

Unfortunately, the accuracy of the velocity reconstruction strongly depends on the number density of tracers, with the velocities of the BGS sample better reconstructed than those of its LRG counterpart. Despite these numerical limitations, the use of the PIZA algorithm provides dynamical information beyond the linear approximation \citep{croft1997reconstruction}. In any case, we would like to stress that results related to the use of the LRG catalogue must be taken with caution due to the inherent technical difficulties just described. Part of this issue was mitigated by applying a radius cut (see Sect.~\ref{s:void_catalogue}) that discards noise-dominated voids.

\subsection{Void finder}
\label{s:voidfinder}

\begin{table}[h!]
\centering
\caption{Brief description of the physical thresholds employed by \texttt{AVISM}.}
\label{tab:thresholds}
\begin{tabular}{c|l}
\hline
Threshold & Description \\
 \hline  
$\delta_1$ & Density contrast threshold for seeding a void \\
$\delta_2$ & Maximum density contrast value inside voids \\
$|\nabla \delta|_\mathrm{th}$ & Maximum density gradient allowed inside a void \\
$\nabla\cdot\mathbf{v}_\mathrm{th}$ & Minimum velocity divergence inside voids  \\
\end{tabular}
\end{table}

For the purpose of this work, we employed the publicly available\footnote{\url{https://github.com/oscarmonllor99/AVISM}} \texttt{AVISM} code \citep{monllor2025avism} to generate the void catalogues from the BGS and LRG samples described in Sect.~\ref{s:survey}. This void finder is able to locate voids within simulations (either for particle or cell data), halo catalogues, and survey (observational) data. We provide a summary of its void-finding methodology below. 

\begin{itemize}
    \item It employs the density, density gradient, and velocity divergence fields to physically define void regions. 
    These quantities are defined on a computational volume discretised using fixed-size cubic cells.
    The shape of the grid can be adapted to the morphology of the working volume depending on the study case. The size of the cubic cells determines the spatial resolution and can also be adapted, depending on the application.
    
    \item When providing particles (galaxies in this study) as input, the data are smoothed on the grid by means of an adaptive smoothing kernel (e.g. \citealt{valles2024vortex, monllor2025avism}), inspired by the smoothed-particle hydrodynamics (SPH) approach. The smoothness of the fields can be controlled by a user-defined parameter, $k$, that sets the number of neighbouring particles that contribute to each cell.

    \item With the physical fields properly defined, the void finder applies four user-defined thresholds in order to carry out the void-finding process (see Table \ref{tab:thresholds} for a brief description of these thresholds). While $\delta_2$ and $|\nabla \delta|_\mathrm{th}$ help in the identification of void edges, the most important threshold correspond to $\delta_1$ and $\nabla\cdot\mathbf{v}_\mathrm{th}$. The first defines which cells have a low enough density to seed a void, whereas the second labels expanding cells.

    \item The use of the velocity divergence is key, as it helps restrict the void-finding process to expanding regions and allows the void finder to prioritise the identification of these structures in cells maximising the expansion, which usually lie in the centre of voids.

    \item Void shapes are unrestricted by the void-finding methodology, as the grid can trace and capture their complex 3D structures. In general, the finer the grid, the better the void finder is at capturing the void shape.

    \item By constructing a hierarchy of nested grids and applying the aforementioned methodology recursively, it is possible to identify voids-in-voids in a simple and consistent manner.
    
\end{itemize}

Among other results, the void finder output provides the centre ($X_c, Y_c, Z_c$), volume ($V$), ellipticity ($\varepsilon$)\footnote{Defined as $\varepsilon = 1 - c/a$, where $a$ and $c$ are, respectively, the lengths of the major and minor axes of the associated ellipsoid.}, inverse porosity ($\mathrm{IP})$\footnote{Defined as $\mathrm{IP} = V_E/V$ \citep{shandarin2006shapes}, where $V_E$ is the volume of the ellipsoid fitting the void and $V$ is the actual void volume.}, mean density, and 3D shape (defined on a grid) for each of the catalogue members. A key feature of this void finder, which must be stated explicitly, is the centring criterion. Instead of considering the volume-weighted mean of cell positions, the void centre is defined as the cell maximising velocity divergence, that is, with maximum expansion.

We incorporated several new features that allow \texttt{AVISM} to process complex observational data such as DESI. These changes are only taken into account if the void finder is intended to work in `survey' mode. The features are listed as follows.

\begin{enumerate}
    \item The code now accepts a mask that restricts the void-finding procedure to those cells of the regular grid that fit within the survey volume.

    \item Survey errors such as fibre assignment incompleteness, imaging systematics, and redshift systematics can be taken into account by providing a weight ($w_\text{tot}$) value for each galaxy.

    \item A random catalogue must be provided with the observational data. These random points are uniformly distributed, capturing the survey window, selection function, and systematics. They also have the weights, $w_\text{tot}$, properly defined to make sure that their weighted (normalised) redshift distribution matches the data (regarding the particular case of DESI DR1, see \citealt{myers2023target} or \citealt{desi2024II} for the definition of weights and random catalogues).

    \item With previous points taken into account, the density contrast for survey data is defined as \citep[e.g. see][for similar definitions of $\delta$]{nadathur2019beyond} 
    \begin{equation}
    \label{eq:delta_gal}
        \delta(\mathbf{x}_\text{cell}) = \frac{D(\mathbf{x}_\text{cell})}{\eta R(\mathbf{x}_\text{cell})} - 1\, ,
    \end{equation}
    with $\mathbf{x}_\text{cell}$ being the position of each cell in the grid, and $D(\mathbf{x}_\text{cell})$ and $R(\mathbf{x}_\text{cell})$ the weighted number of data points and randoms at each cell, respectively. The $\eta$ factor normalises the sum:
    \begin{equation}
        \eta = \frac{\sum_{\text{cells}} w^{(D)}}{\sum_{\text{cells}} w^{(R)}}\, ,
    \end{equation}
    where the sum runs over all cells belonging to the survey mask. The values of $D$ and $R$ for each cell are obtained by distributing galaxy and random (respectively) weighted counts onto a grid using \texttt{AVISM}'s smoothing kernel. 
\end{enumerate}

Both study volumes consist of a cube of $1400 \, \mathrm{cMpc}$ side, extracted from the total BGS and LRG samples (see Sect.~\ref{s:survey}). In both cases, the void finder was applied to a $500^3$ cubic grid, yielding a spatial resolution of $2.8 \, \mathrm{cMpc}$. 

With respect to the void finder parameters, to interpolate the density and velocity fields onto the grid with equal sharpness, the $k$ parameter (number of neighbours that contribute to each cell) must be carefully chosen. While we chose $k = 32$ (the default) for the BGS input, the same parameter had to be modified for the LRG sample to take into account that its number density is $\sim24$ times lower. We found $k = 4$ yielded the best results as a compromise between recovering a sharp density and velocity field and having enough information to interpolate. In particular, we found that $\overline{l}_{32}^\text{\,BGS} \approx \overline{l}_{4}^\text{\,LRG}$, that is, the mean distance from a BGS galaxy to its $32$nd nearest neighbour is roughly the same as the mean distance from an LRG to its $4$th. Thus, by choosing these values of $k$, we preserved the average SPH kernel smoothing length between the two samples, which were thus interpolated with a similar spatial resolution.

With the physical fields similarly defined for both BGS and LRGs, the next step consisted of adjusting the physical thresholds in order to account for the different tracer bias. For both samples, we chose $\nabla \cdot \mathbf{v}_\mathrm{th}= 0$ and $\delta_2 =  10$, which are the default values for these thresholds. Note that $\delta_2$ has little effect on this study due to the inherent smoothness of the interpolated fields because of the low tracer densities, which yield small overdensities. The values of $\delta_1$ and $|\nabla\delta|_\mathrm{th}$ were modified and adapted to each case. For the BGS sample, we set them to $-0.45$
and $0.25 \, \mathrm{cMpc}^{-1}$, respectively. For the LRG case, these values were scaled by a factor of $b_\mathrm{LRG}/b_\mathrm{BGS} \approx 1.7$ in order to take into account the bias effect and, hence, obtain physically compatible void structures across both samples.

\section{Results}
\label{s:results}

In this section, we present the main results obtained after running \texttt{AVISM} on the preprocessed BGS and LRG samples from the DESI DR1 LSS data. In Sect.~\ref{s:void_catalogue} we describe the final catalogues used for the void statistics and profiles provided below, as well as presenting a 2D and 3D visual inspection of different voids. In Sect.~\ref{s:statistics} we provide the void size function along with histograms for the ellipticity, mean density, and inverse porosity. In Sect.~\ref{s:profiles} we display the stacked cumulative density profiles. Finally, in Sect.~\ref{s:bias_in_voids} we estimate the value of the linear bias in BGS and LRG voids.

\subsection{Void catalogues}
\label{s:void_catalogue}

\begin{figure}
\centering 
\includegraphics[width=0.9\linewidth]{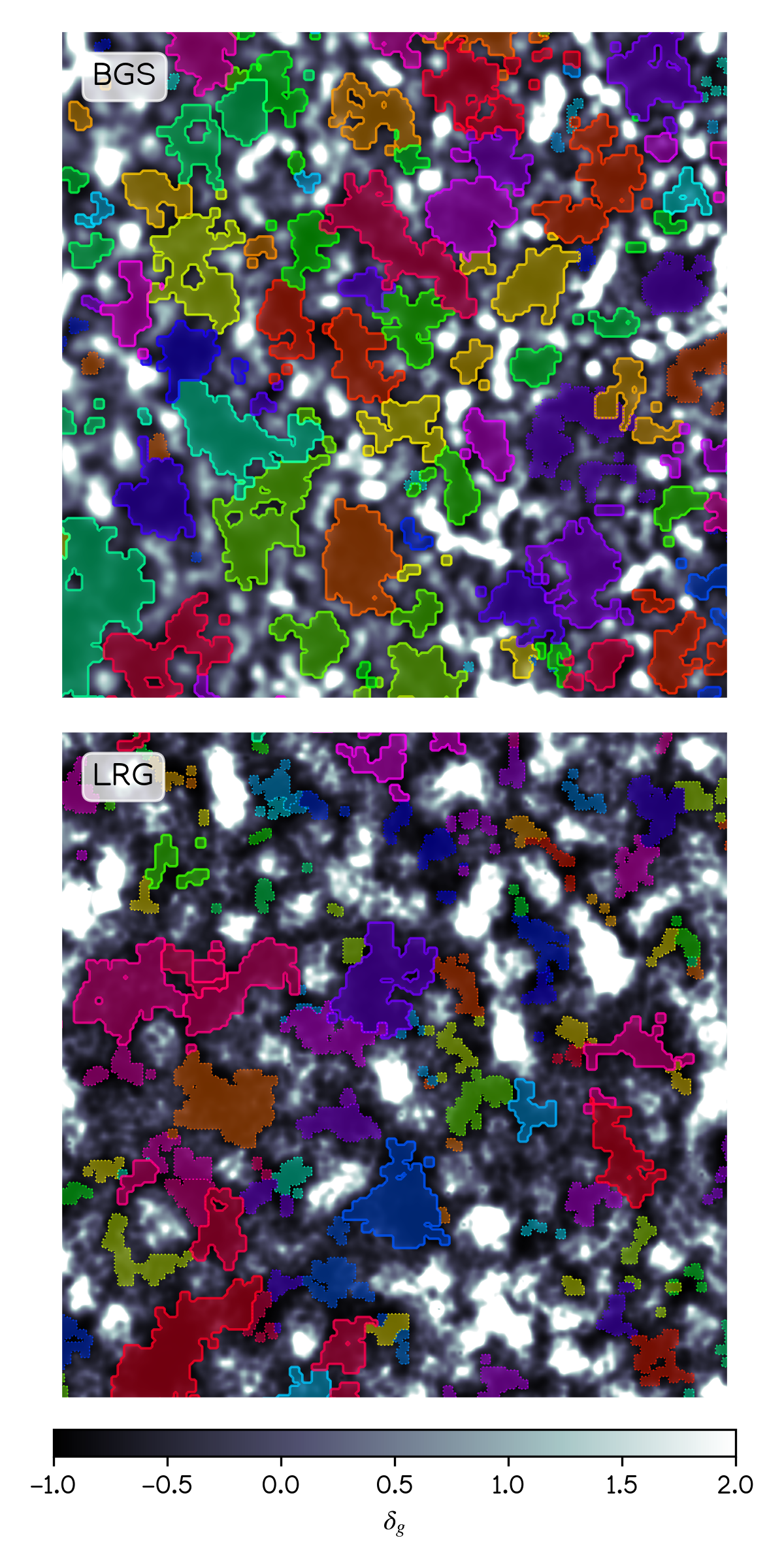}
\caption{Distribution of voids intersecting a thin slice of $500\, \mathrm{cMpc}$ side length both for the BGS (upper panel) and LRG (lower panel) catalogues. Different colours are used to show different void IDs, with solid (dotted) contours displaying accepted (eliminated) voids after the void catalogue cleaning. Voids are shown overlaid on the density contrast field as interpolated by \texttt{AVISM}, represented by a grey colour scale with values displayed in the colour bar below. Only the intersecting part of the void is shown.}
\label{fig:void_visual}
\end{figure}

\begin{figure*}
\sidecaption
\includegraphics[width=10cm]{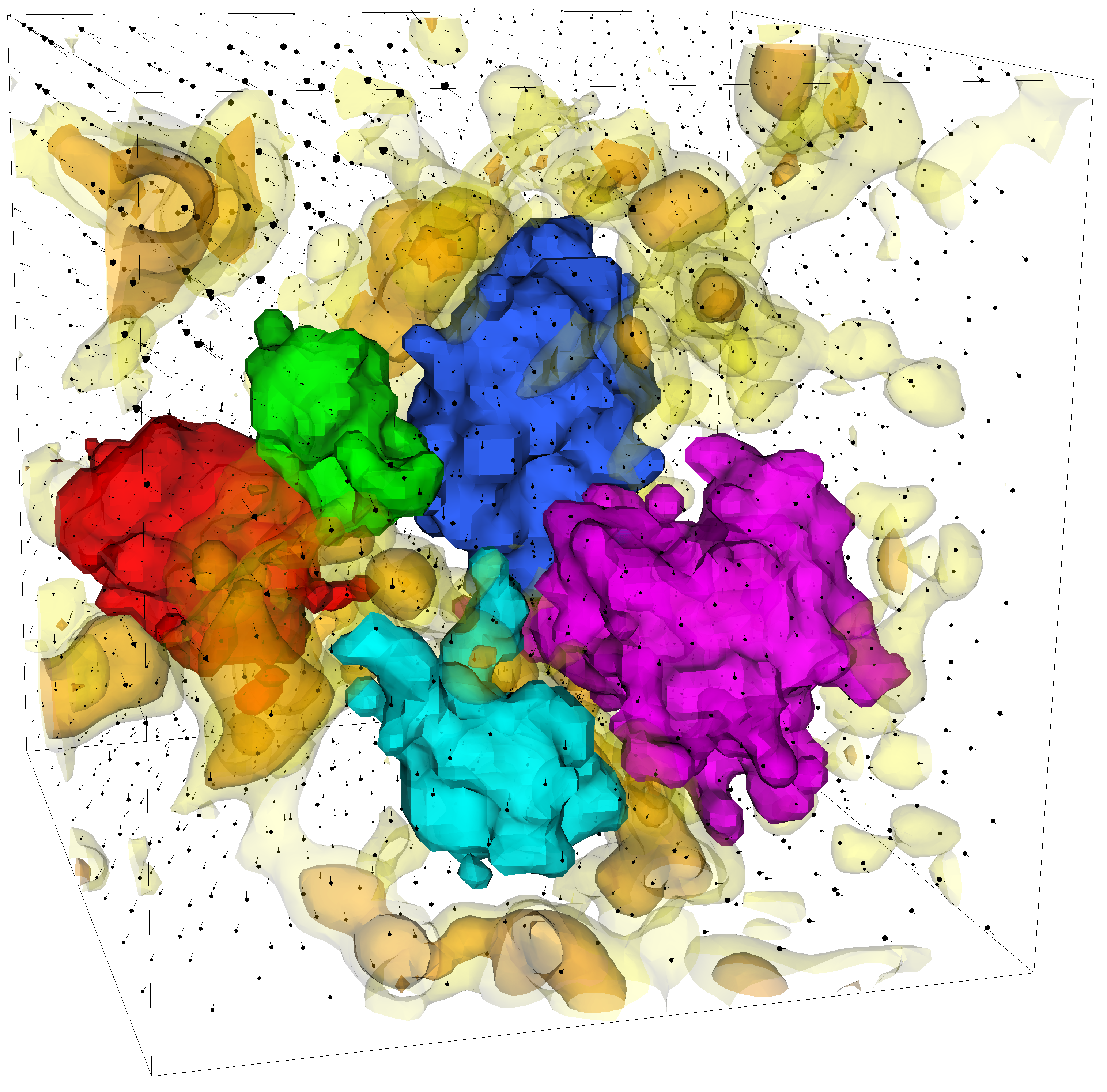}
    \caption{3D contours of the density contrast field (in yellow and orange) together with five voids identified by \texttt{AVISM} depicted in a different colour for each ID. We also represent the velocity field obtained by the PIZA method with black arrows, whose lengths scale with the velocity magnitude. The shown region corresponds to a cubic $200^3 \, \mathrm{cMpc^3}$ sub-volume of the whole BGS survey volume.}
\label{fig:rendering}
\end{figure*}

After running the void finder on both samples, we obtained \mbox{$N_\text{BGS} = 5867$} voids within the BGS data and \mbox{$N_\text{LRG} = 8597$} in the LRG data. Once the void catalogue was extracted, a cleaning step had to be performed to eliminate voids that lie too close to the survey mask boundary. We first identified voids that are just two cells away from the mask boundary, marking them as potential `edge voids'. Then, for these bordering voids, we calculated the fraction of their volume within a distance of two cells to the mask limit. If this fraction was greater than $10\%$, the void was rejected, as we considered that it would have grown significantly larger if more data were available. With this cleaning step, only sufficiently interior voids were accepted and considered for the study, avoiding contamination of the results due to artificial void boundaries. Furthermore, following previous studies \citep{hoyle2002voids,pisani2015counting, cousinou2019multivariate,rincon2025desivast} Poisson noise can dominate the detection of small voids in surveys, and, hence, a radius cut must be applied to leave only void identifications that are significant enough. To obtain a physically motivated cut note first that $R_e = (\frac{3}{4\pi}V)^{1/3}$ is the void effective radius. In parallel, the mean galaxy separation ($\lambda$) can be obtained for each sample as $\lambda = \overline{n}_g^{\,-1/3}$, with $\overline{n}_g$ being the average number density of galaxies inside the sample volume. In order to be physically consistent, the radius cut should be larger than the mean galaxy separation. In this work, we opted for
\begin{equation}
    \label{eq:rad_cut}
    R_e > 1.5\, \lambda \,,
\end{equation}
resulting in $R_e \gtrsim 12 \, \mathrm{cMpc}$ and $R_e \gtrsim 33 \, \mathrm{cMpc}$ for BGS and LRG voids, respectively. Although more conservative radius cuts have been presented in the literature, we adopted a more permissive approach primarily because \texttt{AVISM} is less prone to identifying noise-dominated voids than spherical or watershed void finders. It uses not only the density field, but also the density gradient and velocity divergence to find voids. In particular, if the velocity field is random and not consistent with that of an expanding region, \texttt{AVISM} will not be able to detect a void. 

After applying the edge and radius cut cleaning steps, we had \mbox{$N_\text{BGS} = 1114$} and \mbox{$N_\text{LRG} = 216$} clean voids. Note that significantly fewer clean voids were obtained for LRGs due to their lower number density, which ultimately translates into a larger mean galaxy separation and radius cut.

For a qualitative visual inspection, in Fig.~\ref{fig:void_visual} we display voids intersecting a thin slice of the $500 \, \mathrm{cMpc}$ side for both the BGS and LRG catalogues, with the interpolated density contrasts overlaid. Only the intersecting part of the void is depicted, which explains some unusual void shapes. So as to display the 3D shape of voids identified by \texttt{AVISM}, Fig.~\ref{fig:rendering} shows the 3D contours of the density contrast inside a cubic $200^3 \, \mathrm{cMpc^3}$ sub-volume of the whole BGS survey volume together with five voids identified within this region.

\subsection{Statistics}
\label{s:statistics}

\begin{figure}[h!]
\centering 
\includegraphics[width=0.96\linewidth]{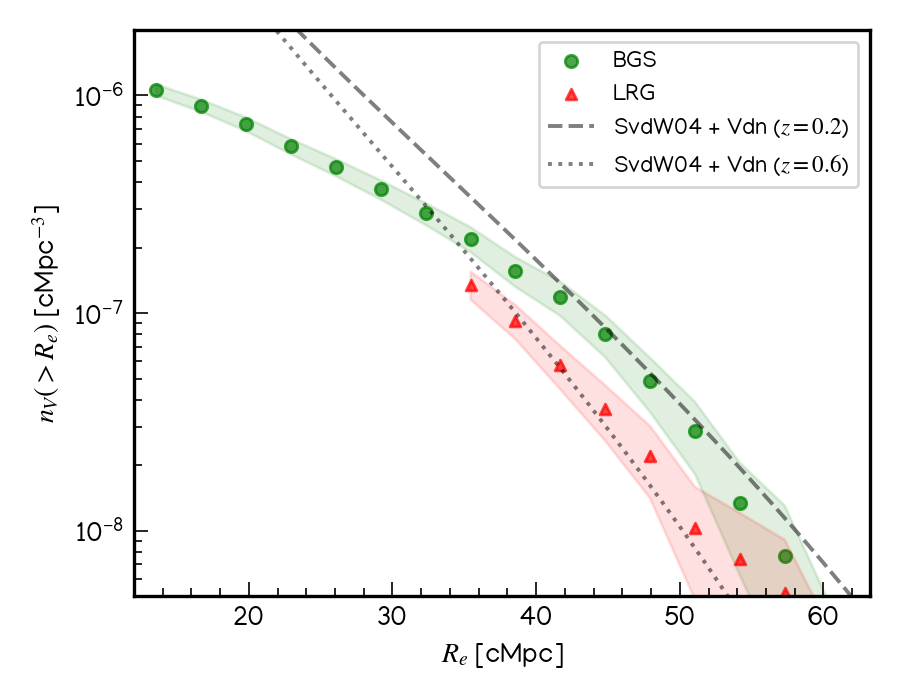}
\caption{Cumulative void size functions from the BGS (green circles) and LRG (red triangles) void catalogues. We also plot the theoretical SvdW+Vdn model \citep{sheth2004hierarchy, jennings2013abundance} at each sample effective redshift (dashed line for $z=0.2$ and dotted line for $z=0.6$) in grey. The linear density contrast threshold is set to $\delta_V^{\mathrm{L}} = -0.55$ in both cases. The shaded region displays $2\sigma$ of the Poisson shot noise error.}
\label{fig:VSF}
\end{figure}

\begin{figure*}[h!]
\centering 
\includegraphics[width=0.9\linewidth]{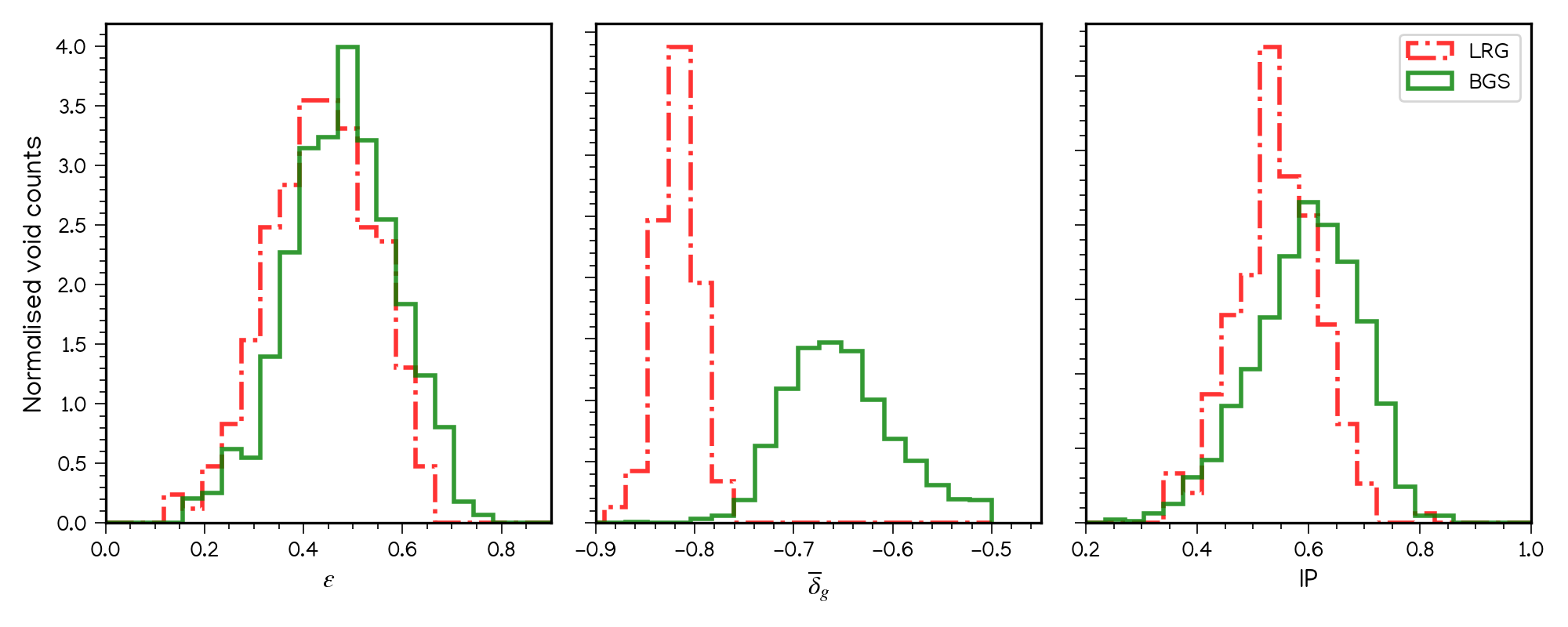}
    \caption{Histograms (void counts) for the ellipticity ($\varepsilon$), mean density contrast ($\overline{\delta}_g$), and inverse porosity (IP) for all BGS (green) and LRG (red) voids in the final (cleaned) void catalogues.}
\label{fig:hists}
\end{figure*}

\begin{figure*}[h!]
\centering 
\includegraphics[width=0.95\linewidth]{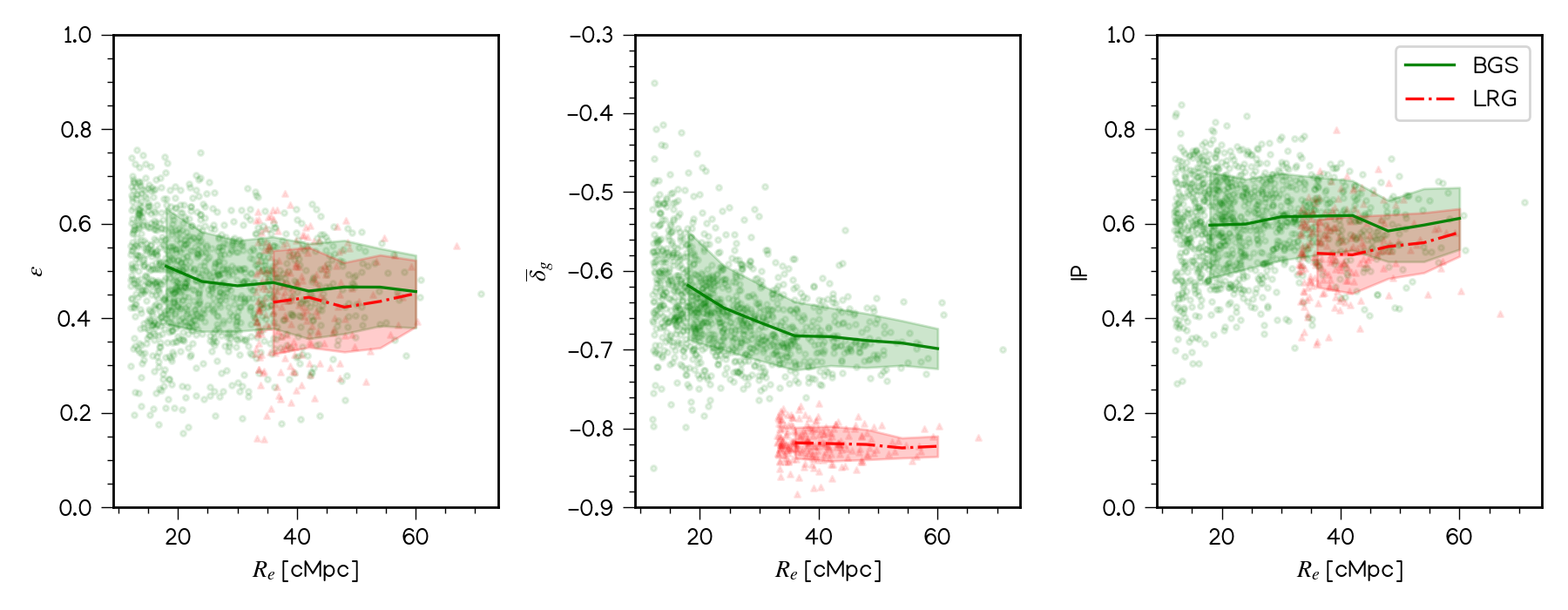}
    \caption{Scaling relations of ellipticity (left), mean density contrast (middle) and inverse porosity (right) as a function of void effective radius ($R_e$) for both the BGS (solid green line and circles) and LRG (dash-dotted red line and triangles) void catalogues. We plot the mean across all voids inside each radial bin, with the shaded region representing one standard deviation.}
\label{fig:stat_rad}
\end{figure*}

The void size function (VSF), that is, the number of voids for a given effective radius (volume), represents one of the most important statistics when studying voids. Since it describes the abundance of underdense regions in the Universe, it makes it possible to constrain the overall matter distribution and its evolution, becoming a key probe of the cosmological model itself \citep{contarini2023cosmological,contarini2024perspective,fernandez2025constraining}. Several theoretical works have been devoted to obtaining an analytical expression for the VSF. In particular, we considered the fit for spherical voids presented in \citet{sheth2004hierarchy} (henceforth the SvdW model) developed under the excursion-set formalism and later refined in \citet{jennings2013abundance} (Vdn model) by considering that the total void volume is conserved in the linear-to-non-linear mapping (hence, it is a volume-conserving model). The key parameter on which the SvdW+Vdn model depends is
\begin{equation}
\mathcal{D} = \frac{|\delta_{V}^{\rm L}|}{\delta_{c}^{\rm L} + |\delta_{V}^{\rm L}|},
\end{equation}
where $\delta_c^{\mathrm{L}} = 1.686$ is the critical overdensity for gravitational collapse (well constrained by the spherical collapse model) and $\delta_V^{\mathrm{L}}$ is the linearly extrapolated density contrast threshold for shell-crossing\footnote{The shell-crossing time is defined as the time when the faster inner shells catch the slower outer ones.}. By varying the $\delta_V^{\mathrm{L}}$ free parameter, one can fit a theoretical VSF to the data, even if voids were identified using the whole mass content or tracers \citep{sutter2014sparse}. It also has to be taken into account that once $\delta_V^{\mathrm{L}}$ is set, the non-linear density contrast threshold for shell-crossing ($\delta_V^{\mathrm{NL}}$) is directly related to it, and it can be obtained by applying the \citet{bernardeau1994} formula:
\begin{equation}
    \label{eq:bernardeau}
    \delta^{\mathrm{NL}} = \bigg(1 - \frac{\delta^{\mathrm{L}}}{C} \bigg)^{-C} - 1\, ,
\end{equation}
where $C = 1.594$. With the value of $\delta_V^{\mathrm{NL}}$, one can close the Vdn model by fixing the linear-to-non-linear size scaling, given by mass conservation:
\begin{equation}
    \label{eq:size_scaling}
    r_V^{\mathrm{NL}} = \frac{r_V^{\mathrm{L}}}{(1+\delta_V^{\mathrm{NL}})^{1/3}}\, ,
\end{equation}
where $r_V^{\mathrm{L}}$ is the void linear radius and $r_V^{\mathrm{NL}}$ is the void radius at the time of shell-crossing. Other models beyond SvdW+Vdn have been presented \citep{verza2024universal}, although for the sake of simplicity, we restricted ourselves to the former in the context of this work. Moreover, rather than a precise comparison between a theoretical and an observational VSF, this section outlines a rough approximation to the problem. The tenet of the SvdW+Vdn model is that voids are spherical and all have the same mean density contrast after shell-crossing. These model conditions do not apply to \texttt{AVISM} voids, as this void finder can handle arbitrary shapes and $\overline{\delta}$ values.

In Fig.~\ref{fig:VSF}, we provide the cumulative VSF both for the BGS (green circles) and LRG (red triangles) void catalogues, together with the theoretical SvdW+Vdn model calculated at each sample's effective redshift ($z_\text{eff}=0.2,0.6$, respectively) using the same value of $\delta_V^{\mathrm{L}} = -0.55$. The plot is restricted to $R_e > 12 \, \mathrm{cMpc}$, as all voids with sizes below this limit were deleted in the cleaning step. Both samples display similar VSF shapes and are compatible with the theoretical fit within 2$\sigma$ of the Poisson error for $R_e \gtrsim 35 \,\mathrm{cMpc}$. The choice of void-finding parameters described in Sect.~\ref{s:voidfinder} has proven successful in identifying physically compatible void populations across both galaxy samples, since they properly follow the theoretical VSF shape and its expected evolution with redshift. For $R_e \lesssim 35 \,\mathrm{cMpc}$, the number of resolution elements, in both galaxy counts and cells, limits the ability to detect small voids in the BGS and LRG volumes. This would explain the gap between the theoretical and observational size functions in this size range.

The VSF is a 1D statistic without information about the 3D structure of voids. To provide insight into this matter, in Fig.~\ref{fig:hists} we display histograms for the ellipticity, mean density contrast, and inverse porosity for both the BGS (solid green line) and LRG (dash-dotted red line) void catalogues. With respect to ellipticity, some studies find that voids are mildly aspherical, with $\varepsilon \sim 0.15$ \citep[e.g.][]{sutter2014sparse, schuster2023cosmic, schuster2024cosmic}, while others find $\varepsilon \sim 0.55$ on average for $0 <z < 2.5$ \citep[e.g.][]{ricciardelli2013structure}. Clearly, the void-finding methodology has an impact, and there is no consensus in the literature. In our case, we obtained a similar distribution of ellipticities for both LRG and BGS samples, with a mean ellipticity of $\varepsilon \sim 0.5$, though BGS voids are slightly shifted towards higher ellipticities. We found a similar scenario for the inverse porosity, which is slightly higher for the BGS. On the other hand, the mean tracer density contrast inside voids clearly increases from LRGs ($\overline{\delta}_g \sim -0.8$) to BGS ($\overline{\delta}_g \sim -0.65$). This seems puzzling; as gravity pulls matter out of voids into the more massive structures surrounding them (clusters, filaments, and walls), one should expect that voids become emptier as time progresses. However, because LRGs are more biased than BGS, they leave larger empty zones between tracers, yielding less dense voids and compensating for the natural evolution towards more pronounced underdensities. To isolate the void depletion process from the bias effect and study its evolution, the cumulative profiles are a promising tool, which we investigate in Sect.~\ref{s:profiles}.

Though Fig.~\ref{fig:hists} already provides insight into the 3D structure of voids, these quantities can vary as a function of void size ($R_e$). To study this dependency, the scaling relation (binned mean) of the same quantities displayed in Fig.~\ref{fig:hists} against the void effective radius is provided in Fig.~\ref{fig:stat_rad} for both the BGS and LRG void catalogues. On average, there is no significant trend in ellipticity or inverse porosity with respect to void size. On the other hand, our results suggest that larger voids are emptier, as the mean density contrast inside BGS voids decreases with increasing $R_e$.

\subsection{Density profiles}
\label{s:profiles}

\begin{figure}[h!]
\centering 
\includegraphics[width=0.94\linewidth]{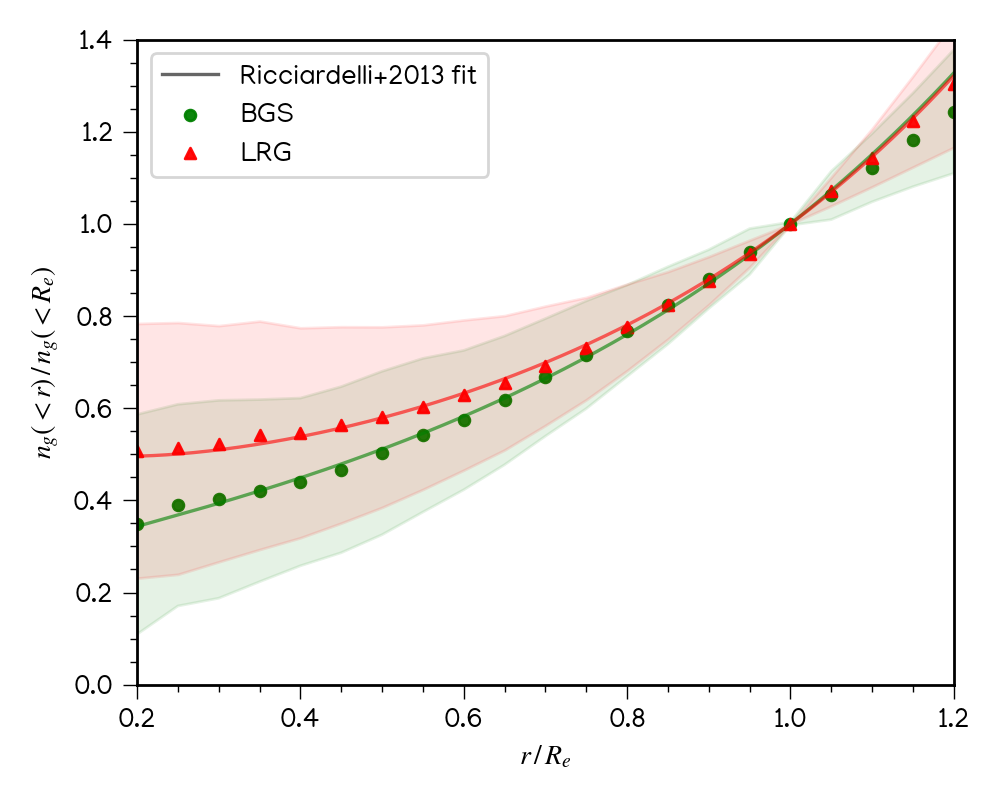}
\caption{Mean cumulative density radial profile for the BGS (green circles) and LRG (red triangles) void catalogues. The shaded region displays one standard deviation. In solid lines, the \citet{ricciardelli2013structure} empirical fit is provided for both void samples.}
\label{fig:profiles}
\end{figure}

\begin{figure}[h!]
\centering 
\includegraphics[width=0.94\linewidth]{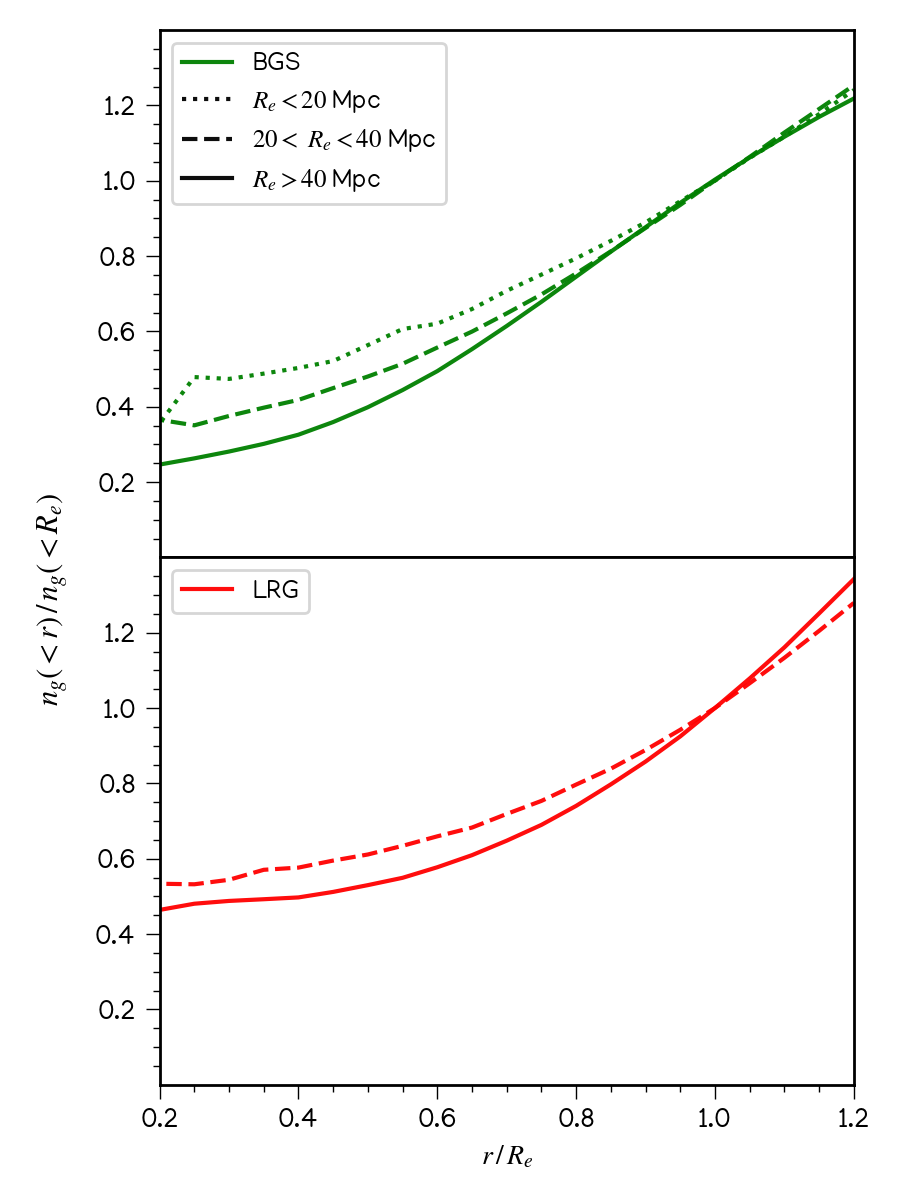}
    \caption{Same as Fig.~\ref{fig:profiles} but carrying out the average over three different size bins. BGS (LRG) profiles are displayed in the top (bottom) panel. For simplicity, standard deviations were deliberately omitted.}
\label{fig:profiles_segregate}
\end{figure}

In the previous section, we analysed the BGS and LRG catalogues from a purely statistical point of view. Now, we focus on the void density profiles, displaying the spatially resolved matter (tracer) distribution inside voids.

A key finding from both observational and cosmological simulation works is that voids display a universal spherical density profile when the distance to the void centre is scaled to $r/R_e$ and a sufficient number of voids is used to stack the variety of profiles. The first efforts focused on properly describing the inner part \citep[$r/R_e \lesssim1$; e.g.][]{van1993voids,colberg2005voids, padilla2005spatial,ricciardelli2013structure, ricciardelli2014universality}, while later works \citep[e.g.][]{sutter2012public,hamaus2014universal, nadathur2014universal} also studied the wall part, beyond $r = R_e$. For this particular work, we restricted our analysis to $r \lesssim R_e$, since we do not expect to properly reproduce the wall prescriptions provided by previous studies due to the low spatial resolution ($\approx 3 \, \mathrm{cMpc}$) needed to accommodate the relatively low number density of matter tracers on the grid. In this regard, we took the empirical cumulative profile presented in \citet{ricciardelli2013structure},
\begin{equation}
    \label{eq:ricciar}
    \frac{\rho(<r)}{\rho(< R_e)} =  \bigg(\frac{r}{R_e}\bigg)^\alpha \exp\left(\bigg[\frac{r}{R_e}\bigg]^\beta - 1\right)\,\, ,
\end{equation}
and fitted the $\alpha$ and $\beta$ values for each study case, hence checking if BGS and LRG voids follow the universal density profile obtained by previous theoretical and numerical studies.

We calculated\footnote{Note that we omitted $r/R_e < 0.2$ on purpose, as only a small number of cells contribute to the profile at those distances.} the cumulative radial number density ($n_g$) profile for each void by using the galaxy density contrast ($\delta_g$) obtained by applying Eq.~\eqref{eq:delta_gal} and normalised it by the mean density inside $R_e$. The maximum velocity divergence point (from which the void identification starts, see \citealt{monllor2025avism}) was chosen as the centre. The results are displayed in Fig.~\ref{fig:profiles}, which shows the normalised void density profiles for BGS (green circles) and LRG (red triangles) samples. The corresponding best fits using Eq.~\eqref{eq:ricciar} are displayed as solid lines. Both samples are well described by this analytical profile, which was previously constrained using cosmological simulations and observations. As a general trend, the LRG profile is shallower than the BGS one. Also, the former lies above the latter within $R_e$. We note that, although in absolute terms the density in LRG voids is lower because of the bias effect, we circumvented such an issue here as these are relative profiles normalised to the respective mean density inside the effective radius. Hence, BGS underdensities are relatively deeper than those of LRGs. This result is in agreement with previous theoretical \citep[e.g.][]{sheth2004hierarchy} and numerical \citep[e.g.][]{ricciardelli2013structure} works; that is, as time progresses, matter flows outwards from voids, making them emptier and continuously increasing their radial density gradient.

Regarding how void profiles may change for different sizes, Fig.~\ref{fig:profiles_segregate} shows the mean cumulative radial number density profiles for three broad groups of void sizes --- small ($R_e < 20 \, \mathrm{cMpc}$), medium ($20 < R_e < 40 \, \mathrm{cMpc}$), and large ($R_e > 40 \, \mathrm{cMpc}$) --- for the BGS sample; medium and large sizes are shown for the LRG sample. As expected, in the inner part, the density profiles show lower central density and steeper gradients as the void size increases.
Interpreting this through the \citet{sheth2004hierarchy} spherical void model (where initial underdensities naturally evolve into deeply depleted, steep-edged profiles over cosmic time), these findings indicate that the smaller voids in our sample are at an earlier stage of mass depletion.

\subsection{The value of bias in DESI voids}
\label{s:bias_in_voids}

\begin{figure}[h!]
\centering 
\includegraphics[width=1\linewidth]{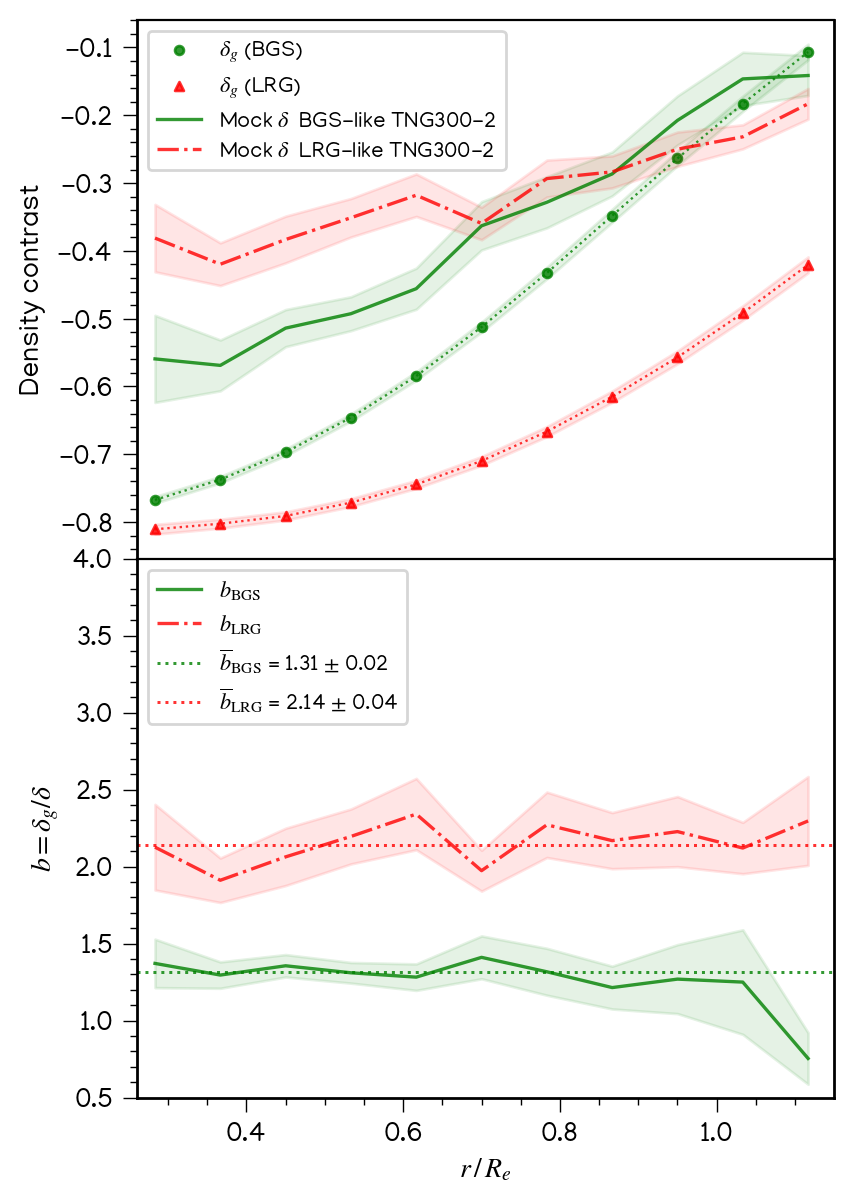}
\caption{Upper panel: Mean differential density contrast profiles for the BGS (green circles) and LRG (red triangles) voids together with the $z \approx 0.2$ (solid green line) and $z \approx 0.6$ (dash-dotted red line) mean $\delta$ profiles obtained using all dark matter particles from a cosmological simulation. Lower panel: Radial profile of the linear bias defined as $b = \delta_g / \delta$, both for the BGS (solid green line) and LRGs (dash-dotted red line). The shaded regions depict the $1\sigma$ standard error. The mean linear bias values of BGS and LRGs (calculated within $r<R_e$) are displayed as horizontal lines for reference.}
\label{fig:diff_profiles_bias}
\end{figure}

The concept of bias (see Sect.~\ref{s:gal_bias}) is rooted in how the distribution of tracers is related to the underlying distribution of matter. At large scales, one can simply use Eq.~\eqref{eq:bias} and calculate the linear bias $b$ by measuring the 2PCF or power spectrum of the distribution of tracers relative to the whole matter content (see Sect.~\ref{s:gal_bias}). However, it is not straightforward to apply this idea to underdense regions. Previous studies carried out in and around void environments led to the conclusion that a simple constant linear bias relation exists in voids, although the value of bias varies with respect to the LSS measurements: $b_\text{voids} \gtrsim b$ \citep{contarini2019cosmological}. Nonetheless, studies following a similar methodology \citep[e.g.][]{verza2022halo} and others following individual galaxy bias estimates \citep[e.g.][]{montero2025galaxy} find that not only is the bias not spatially constant, it increases outwards.

Our goal in this section is to estimate the bias of BGS and LRGs inside voids. Previously mentioned works had the advantage of knowing the full matter content, as they used cosmological simulations. Unfortunately, in our approach we do not know the underlying matter distribution in DESI, as only a limited number of biased tracers are provided. To overcome this important issue, we propose the methodology steps listed below. 

\begin{enumerate}
        \item Extract mock ad hoc galaxy samples from a cosmological simulation mimicking the redshift and clustering (bias) properties of BGS and LRGs. To match the clustering linear bias of BGS and LRGs, follow the \cite{tinker2010large} prescriptions and apply the following mass cuts to the galaxy samples: $M_\mathrm{200m} > 3.8 \times10^{12} \, M_\odot$ and $M_\mathrm{200m} > 1.2 \times10^{13} \, M_\odot$ for the $z \approx 0.2$ and $z \approx 0.6$ snapshots, respectively.

        \item Run \texttt{AVISM} with equivalent spatial resolution, smoothing ($k$), and physical thresholds to the BGS and LRG runs on these mock catalogues. Apply the same radius cut criteria (Eq.~\ref{eq:rad_cut}).

        \item Test and verify the compatibility of the number density profiles from voids obtained using observational and mock void galaxy catalogues.

        \item With verified compatibility, the mass profile of voids identified in the mock galaxy catalogues can be used together with the number density profile of observational voids to estimate the linear bias for $r<R_e$.
\end{enumerate}

For the extraction of the mock galaxy catalogues, we considered the \texttt{TNG300$-$2} cosmological simulation from the \texttt{IllustrisTNG} suite \citep{nelson2019illustristng}. The simulation evolves a cubic volume of $302.6\;\mathrm{cMpc}$ side length, containing $1250^3$ gas and dark matter particles from redshift $z = 127$ to $z = 0$. It incorporates a comprehensive galaxy formation model that accurately tracks the formation and evolution of galaxies over cosmic time \citep{weinberger2016simulating, pillepich2018simulating}. This simulation uses a set of values for the cosmological parameters compatible with our fiducial cosmology (see Sect.~\ref{s:fid_cosm}). Unlike the survey volume, the simulation box is periodic and unmasked. Following the methodology described above, we found 29 and 34 BGS-like and LRG-like clean (above the radius cut) mock voids, respectively.

The compatibility between the observational and mock number density profiles is discussed and established in Appendix~\ref{app:obs_mock_equivalence}. For both the BGS and LRG cases, the relative error lies below 10\% for $r<R_e$. This fact proves the similarity between both (observational and mock) void populations inside this region, therefore justifying the use of the mock matter profiles to estimate the linear bias inside the observational DESI voids.

With previous considerations in mind, in the top panel of Fig.~\ref{fig:diff_profiles_bias} we display the mean number density profile of BGS and LRG voids, together with the corresponding mean matter profiles of the voids identified by applying \texttt{AVISM} on the mock BGS-like and LRG-like samples from the \texttt{TNG300-2} described above. In the lower panel, we display the value of the linear bias defined by Eq.~\eqref{eq:bias} as a function of the void-centric distance ($r$) in units of $R_e$. In agreement with previous theoretical studies \citep{pollina2017linearity}, we found that the value of the linear bias inside voids is roughly constant for $r<R_e$ and slightly higher than the clustering (LSS) value \citep[see][for example, for a similar result with simulations]{contarini2019cosmological}. In particular, we found that the mean values inside $r<R_e$ correspond to $\overline{b}_\mathrm{BGS} = 1.31 \pm 0.02$ and $\overline{b}_\mathrm{LRG} = 2.14 \pm 0.04$, which are, in both cases, mildly larger than the clustering (LSS) counterparts presented in Sect.~\ref{s:gal_bias}.

For the sake of completeness, Fig.~\ref{fig:diff_profiles_bias} shows that the number density profile for LRG voids does not reach $\delta_g \approx 0$ within $r<1.2R_e$. This would point to a possible limitation of \texttt{AVISM} to properly identify void edges in LRG data. This issue could be overcome in future work with a higher number density sample, which would provide a better description of the density and velocity fields underpinning void identification.

\section{Summary and conclusions}
\label{s:conclusions}

The main goal of this work was to provide new insight into and highlight constraints of the structure and evolution of cosmic voids when analysing observational data from a state-of-the-art survey such as DESI. We obtained two void catalogues after processing DESI DR1 BGS and LRG samples with the \texttt{AVISM} void finder. After cleaning edge and noise-dominated voids, we had \mbox{$N_\text{BGS} = 1114$} and \mbox{$N_\text{LRG} = 216$} voids within the BGS and LRG samples, respectively, with sizes up to $\sim 70 \, \mathrm{cMpc}$. Before running the algorithm, a preprocessing pipeline consisting of a two-step redshift-space distortion correction and a velocity reconstruction was applied to the data, the elements of which are listed below:

\begin{itemize}
    \item Fingers of God were compressed along the line of sight to spherical objects after applying the friends-of-friends algorithm with a linking length that is larger in the direction parallel to the line of sight.

    \item Linear redshift-space distortions (Kaiser effect) were corrected with the standard methods used for BAO analysis. In particular, we used the \texttt{pyrecon} package implementing the \citet{burden2015reconstruction} approach.

    \item The accuracy of the corrections (especially regarding the FoG case) was addressed by calculating the two-dimensional, two-point correlation function before and after correcting the data. In both cases, the multipole moments beyond the monopole were reduced, though our methodology performed significantly better with the BGS than the LRG sample.  

    \item The underlying velocity field was reconstructed under the Lagrangian formalism and Zel'dovich's approximation \citep{zel1970gravitational} by applying the iterative PIZA method \citep{croft1997reconstruction} with a series of upgrades to accelerate convergence. The mildly non-linear velocity field reconstructed with this methodology allows \texttt{AVISM} to exploit both the geometrical and dynamical information to carry out a more complete and robust void-finding process.
\end{itemize}

With the obtained BGS and LRG void catalogues, we carried out several analyses to uncover the structure and evolution of voids. These are listed below.

\begin{enumerate}

    \item The cumulative VSFs of BGS and LRG voids are comparable to the theoretical VSF prescription (SvdW+Vdn model) at both effective redshifts. This compatibility also supports the evolution of the theoretical VSF with redshift.

    \item The structure of voids was characterised by computing their ellipticities, mean densities, and inverse porosity. Both BGS and LRG voids have, on average, roughly similar ellipticities ($\varepsilon \sim 0.5$), with the first being slightly more elliptical and less porous. On the other hand, LRG voids are emptier due to their higher bias. We also found that the void mean density decreases with increasing void size, while the ellipticity and inverse porosity remain constant under variations of $R_e$.

    \item The cumulative number density profiles for all voids are compatible with a universal density profile, as presented in previous studies \citep{colberg2005voids, ricciardelli2013structure, hamaus2014universal}. When the cumulative profiles are normalised by their void mean density, BGS voids present a deeper and steeper profile compared with the LRG one, which is in line with the theoretical evolution of voids \citep{sheth2004hierarchy}. When analysing the density profiles for several void size groups, smaller voids are less empty than larger voids, and they also exhibit a shallower profile, which is compatible with an earlier state in their evolution.

    \item Galaxy bias in voids is still a matter of debate. Using the mean number density profile from the two void catalogues presented in this paper, and comparing with the matter density profiles of voids identified within two ad hoc mock galaxy samples (extracted from the \texttt{TNG300-2} cosmological simulation mimicking the properties of the BGS and LRG samples), we estimated this quantity. The results show that $b$ is roughly constant within $r<R_e$, with a mean value of $\overline{b}_\mathrm{BGS} = 1.31 \pm 0.02$ and $\overline{b}_\mathrm{LRG} = 2.14 \pm 0.04$, which in both cases is slightly higher than the clustering (LSS) counterpart ($b_\mathrm{BGS} = 1.22 \pm 0.06$ and $b_\mathrm{LRG} = 2.03 \pm 0.11$), as already pointed out by theoretical works \citep{contarini2019cosmological}.
    
\end{enumerate}

\section{Data availability}
\label{s:data_availability}

For reproducibility and to contribute to previous public observational void catalogues (see the appendix of \citealt{contarini2026era}), we publicly release both the BGS and LRG ASCII catalogues in a GitHub repository: \url{https://github.com/oscarmonllor99/AVISM_BGS_LRG_voids}. On the contrary, due to their large size (several GB), the data cubes describing the 3D structure and physical variables (density and velocity) will be provided upon request.

\begin{acknowledgements}

   We thank the anonymous referee for the valuable insight
and constructive comments that helped improve the quality of this manuscript. This work has been supported by the Agencia Estatal de Investigación Española (AEI; grant PID2022-138855NB-C33) and by the Generalitat Valenciana (grant PROMETEO CIPROM/2022/49). OM acknowledges support from Universitat de València through the Atracció de Talent fellowship.
     
     This research has made use of the following open-source
     packages: NumPy \citep{harris2020array}, SciPy \citep{virtanen2020scipy}, Matplotlib
     \citep{Hunter:2007}, VisIt \citep{HPV:VisIt}, COLOSSUS \citep{diemer2018colossus}, Astropy \citep{astropy:2013,astropy:2018,astropy:2022} and CAMB \citep{lewis2011camb}.
     
\end{acknowledgements}

\bibliographystyle{aa}
\bibliography{aa61712-26}
\vspace{1 cm}

\begin{appendix}

\section{RSD and AP impact on voids}
\label{app:RSD_AP_impact}

\begin{figure*}[t!]
\centering 
\includegraphics[width=0.95\linewidth]{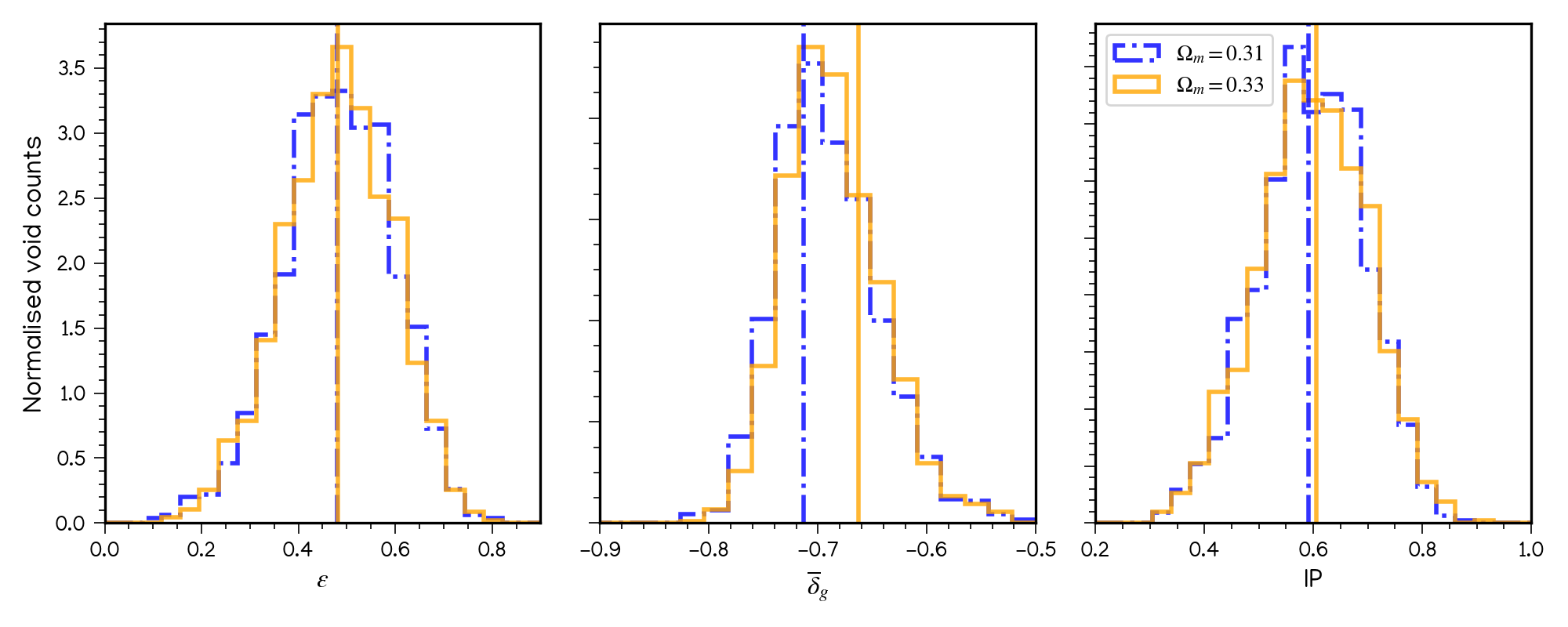}
\caption{Normalised void counts of the ellipticity, mean density contrast and inverse porosity for the BGS data processed with \mbox{$\Omega_\mathrm{m,0} = 0.31$ (blue) and $\Omega_\mathrm{m,0} = 0.33$} (gold) fiducial choices of cosmology and RSDs corrected. 
Vertical lines depict the mean values of the quantities across all voids in the sample. Differences between the mean values are magnified by a factor of 5 to better appreciate the effect.}
\label{fig:AP_impact}
\end{figure*}

\begin{figure*}[t!]
\centering 
\includegraphics[width=0.95\linewidth]{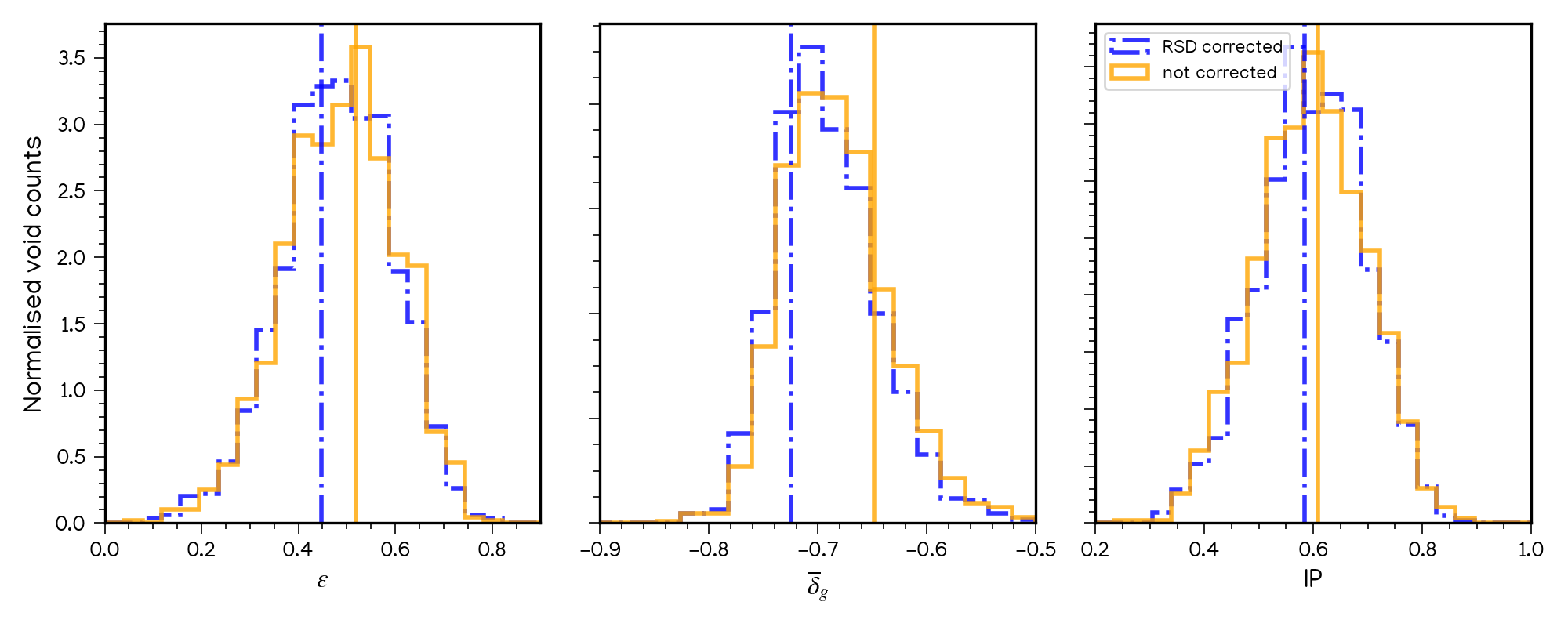}
\caption{Similar to Fig.~\ref{fig:AP_impact} by keeping $\Omega_\mathrm{m,0} = 0.31$ but with RSDs corrected (blue) and untreated (gold).}
\label{fig:RSD_impact}
\end{figure*}

In this appendix, we investigated the impact of both RSDs and the AP effect on the final void catalogue (for the void-finding methodology, see Sect.~\ref{s:voidfinder}). In particular, we quantified to what extent, first, removing the RSD treatment and, later, modifying the $\Omega_\mathrm{m,0}$ parameter from our fiducial $0.31$ to $0.33$ (to test the impact of the AP effect) modified the ellipticity, mean density and inverse porosity of the void statistics found by \texttt{AVISM} when applied on the BGS sample. The results are displayed in Figs.~\ref{fig:AP_impact} and \ref{fig:RSD_impact}. In both cases, the blue histogram corresponds to the statistics found with our fiducial cosmology and correcting RSDs. The gold histogram in Fig.~\ref{fig:AP_impact} corresponds to the results obtained with $\Omega_\mathrm{m,0} = 0.33$ and RSDs corrected. On the other hand, in Fig.~\ref{fig:RSD_impact} the gold histograms were obtained with our fiducial choice of $\Omega_\mathrm{m,0} = 0.31$ but leaving RSDs untreated. In all cases, vertical lines display the mean value of the corresponding quantity. The differences between the blue and gold means are magnified by a factor of 5 for the sake of clarity.

From these figures, we can clearly observe that, under a moderate AP perturbation corresponding to an $\Omega_\mathrm{m,0}$ value substantially discrepant ($5\sigma$) with recent constraints (see \citealt{adame2025desi})\footnote{BAO (DESI DR1) + CMB (Planck) + lensing (Planck + ACT).}, the impact on void properties is smaller than that produced by RSDs. Therefore, for this study, we can safely consider the AP effect as a subdominant effect relative to RSDs.

Complementary to previous points, the impact of FoG and RSDs in general can be quantified in the $\varepsilon$ and $\overline{\delta}_g$ histograms, where the mean ellipticity and density inside voids without correcting RSDs are in both cases $\approx 0.02$ above the mean value for voids found in the corrected data. Although this difference can be considered small, it is statistically significant, given the large void sample $N_\mathrm{BGS} > 1000$, which lowers the standard error to $\sim10^{-3}$ for both quantities.

\section{PIZA method and test}
\label{app:vel_recon_test}

\subsection{Background and implementation}

Under the Lagrangian formalism \citep{peebles1989tracing, buchert1992lagrangian,bernardeau2002large} one can define the displacement field $\mathbf{\Psi}$ which maps the initial (Lagrangian) particle positions $\mathbf{q}$ into the final (Eulerian) particle positions $\mathbf{x}$:
\begin{equation}
    \label{eq:displacement}
    \mathbf{x}(t) = \mathbf{q} +  \mathbf{\Psi} (\mathbf{q}, t)\, ,  
\end{equation}
for any given final time $t$. On sufficiently large scales, the Zel'dovich approximation \citep[ZA;][]{zel1970gravitational} applies and the displacement field is given by \mbox{$\mathbf{\Psi}(\mathbf{q}, t) = D(t) \mathbf{\tilde{\Psi}}(\mathbf{q})$}, where $D(t)$ is the linear growth factor and $\mathbf{\tilde{\Psi}}(\mathbf{q})$ is an irrotational time-independent function of the initial Lagrangian position. Given these conditions, the relation between the displacement field and the initial density field reduces to:
\begin{equation}
    \label{eq:zeldovich_aprox}
    \nabla_{\mathbf{q}} \cdot [\mathbf{\tilde{\Psi}}(\mathbf{q})] = - \delta(\mathbf{q})\, , 
\end{equation}
which is equivalent to the Eulerian Eq.~\eqref{eq:linear} if we multiply by $D(t)$ on both sides. Indeed, the peculiar velocity field is directly related to the displacement field in linear theory by:
\begin{equation}
    \label{eq:velocity_ZA}
    \mathbf{v}(\mathbf{x},t) = f a H \mathbf{\Psi}(\mathbf{q},t)\, ,
\end{equation}
where $\mathbf{x}$ and $\mathbf{q}$ are related by Eq.~\eqref{eq:displacement}. 

Although the connection between Eqs.~\eqref{eq:linear} and \eqref{eq:zeldovich_aprox} through \eqref{eq:velocity_ZA} is evident by taking into account that $\nabla_{\mathbf{q}} \approx \nabla_{\mathbf{x}}$ (which is true at first order), this fact does not imply that both ways of obtaining the velocity field yield equal results. Given $\delta(\mathbf{x})$ (e.g. obtained by smoothing galaxy positions at $z\approx 0$ onto a grid), one can directly apply Eq.~\eqref{eq:linear} and force the linearity to obtain the divergence of the velocity field. Nevertheless, unless the density is sufficiently smooth ($|\delta| \ll 1$, which is the condition for applying Eulerian perturbation theory) this technique will fail to obtain realistic values of the velocity. On the other hand, when using the ZA and Eq.~\eqref{eq:velocity_ZA}, the final (observed) positions of galaxies remain unchanged (they are not `smoothed'), and it is the displacement field that is linearised\footnote{Its perturbative ansatz consists in $t < t_c$ with $t_c$ the time of shell-crossing.}. In this regard, ZA can describe mildly non-linear density contrasts (like the formation of caustics and pancakes). In fact, the Eulerian linear perturbation theory is a limiting case of the ZA \citep{buchert1995lagrangian, hidding2014zel}. This fundamental difference is what allows the ZA to obtain velocities that are accurate even for the mildly non-linear regime.

\paragraph{Solving for $\mathbf{\Psi}$:} given a set of particles (e.g. galaxies in a survey) from which we know the final Eulerian positions $\{\mathbf{x}_i \}_{i=1}^N$, the PIZA approach uses the principle of least action to obtain the particle displacement $\{\mathbf{\Psi}_i\}_{i=1}^N$ that brought them from the initial to their final position. The set of initial positions $\{\mathbf{q}_i \}_{i=1}^N$ is uniformly distributed in space, as it must fulfil the fact that the density field is homogeneous at $t=0$. Then, under the ZA, \citet{croft1997reconstruction} showed that the total action, $S$, is related to the displacements by:
\begin{equation}
    \label{eq:action}
    S   = F(t)\Psi_0 = F(t) \sum_{i=1}^N m_i|\mathbf{\Psi}_i|^2\, ,
\end{equation}
where $\Psi_0$ is the mass-weighted sum of all squared displacements, $m_i$ the mass assigned to each particle (in this case equal for all galaxies), and $F(t)$ is a universal function of time fixed by the cosmology. Hence, according to the principle of least action, the real configuration requires the cost function $\Psi_0$ to be minimised. Solving the variational problem exactly, i.e. finding the global minimum of $\Psi_0$, is complex\footnote{More precisely, the time scaling for obtaining the global solution can be as bad as $\mathcal{O}(N^3)$, which becomes prohibitive for the number of galaxies in DESI.} \citep{frisch2002reconstruction, brenier2003reconstruction}. This is the reason why we opted for the approximate iterative PIZA method. It proceeds as follows:
\begin{enumerate}
    \item Define a set of random initial positions uniformly distributed in space in line with the homogeneity of the Universe at early times.
    \item Randomly assign one of the initial positions to the final position of each particle.
    \item Iterate over random pairs of particles and interchange their initial positions. If $\Psi_0$ is reduced, the interchange is kept and otherwise reversed.
    \item The iterative process runs until the rate of change of $\Psi_0$ falls below a given threshold.
\end{enumerate}

Some improvements over this basic PIZA methodology were implemented in order to, first, accelerate convergence and, second, avoid getting trapped in local minima (it is well known that PIZA can converge to different solutions; \citealt{valentine2000reconstructing}):

\begin{itemize}

    \item Before iterating, instead of randomly mapping the initial positions to the final, we look for the closest 128 initial position neighbours of each galaxy and randomly assign one of the initial neighbours to it. Only $\sim90\%$ of initial positions get assigned with this method within a reasonable time. The rest of the initial and final positions are randomly assigned without taking into account proximity. With this methodology, the iterative process begins from a better starting point. 
        
    \item We also accelerate convergence by restricting initial position interchanges only between close neighbours $\approx66\%$ of the time (probabilistically). On the other hand, proper exploration of swap space is ensured by allowing global swaps to happen with a  $\approx33\%$ probability. The number of neighbours can be tuned to accelerate convergence. We found 256 is a proper value for both the particular BGS and LRG samples. 

    \item To minimise the effect of getting trapped in local minima, instead of using simulated annealing \citep{kirkpatrick1983optimization}, and inspired by \citet{croft1997reconstruction} and \citet{sartori2026back}, we create several sets of random initial positions and run the algorithm in parallel, obtaining different solutions for the displacement field. The average over all random realisations is taken, hence obtaining a more significant final displacement field. 
    
\end{itemize}

Convergence is tracked by checking the relative difference in $\Psi_0$ between iterations. If the relative difference goes below $10^{-4}$ more than 20 times, we consider the algorithm has converged. After that, the displacement field ($\mathbf{\Psi}_i$) is retrieved for all galaxies. Since galaxies are a biased matter tracer, Eq.~\eqref{eq:velocity_ZA} cannot be applied directly. Instead, given the LSS linear bias factor ($b$, see Sect.~\ref{s:gal_bias}), the result of Eq.~\eqref{eq:velocity_ZA} must be multiplied by $1/b$, finally obtaining a realistic, accurate (beyond the Eulerian linear regime) description of the velocity field at each galaxy position. For an explicit derivation of the relation between the biased and total displacement fields, see \citet{sartori2026back}.

\begin{figure}[h!]
\centering 
\includegraphics[width=0.85\linewidth]{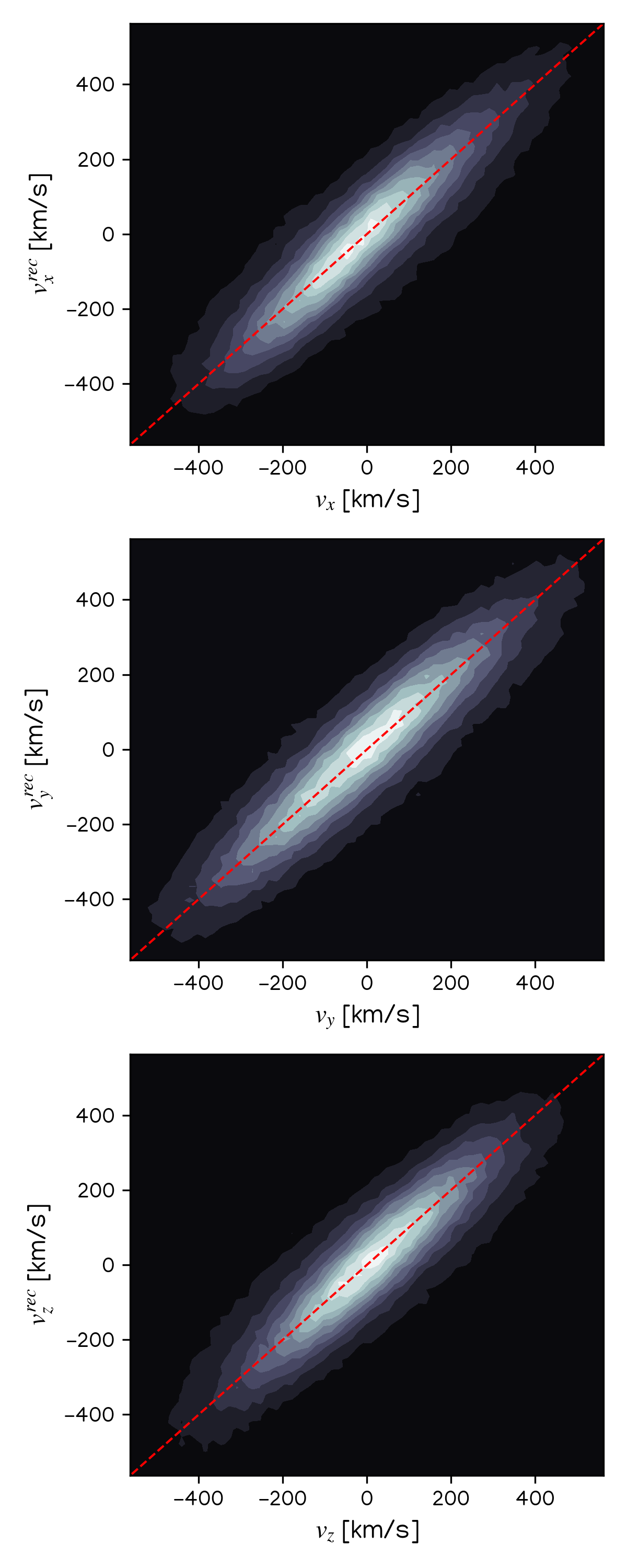}
\caption{Contours of the reconstructed (PIZA) halo peculiar velocity along the $X,Y,Z$ directions as a function of the real (known) peculiar velocity. The dashed red line shows the ideal 1:1 relation. These results correspond to the maximum number density displayed in Fig.~\ref{fig:Pearson_coeff}.}
\label{fig:vel_recon_test}
\end{figure}

\begin{figure}[h!]
\centering 
\includegraphics[width=0.85\linewidth]{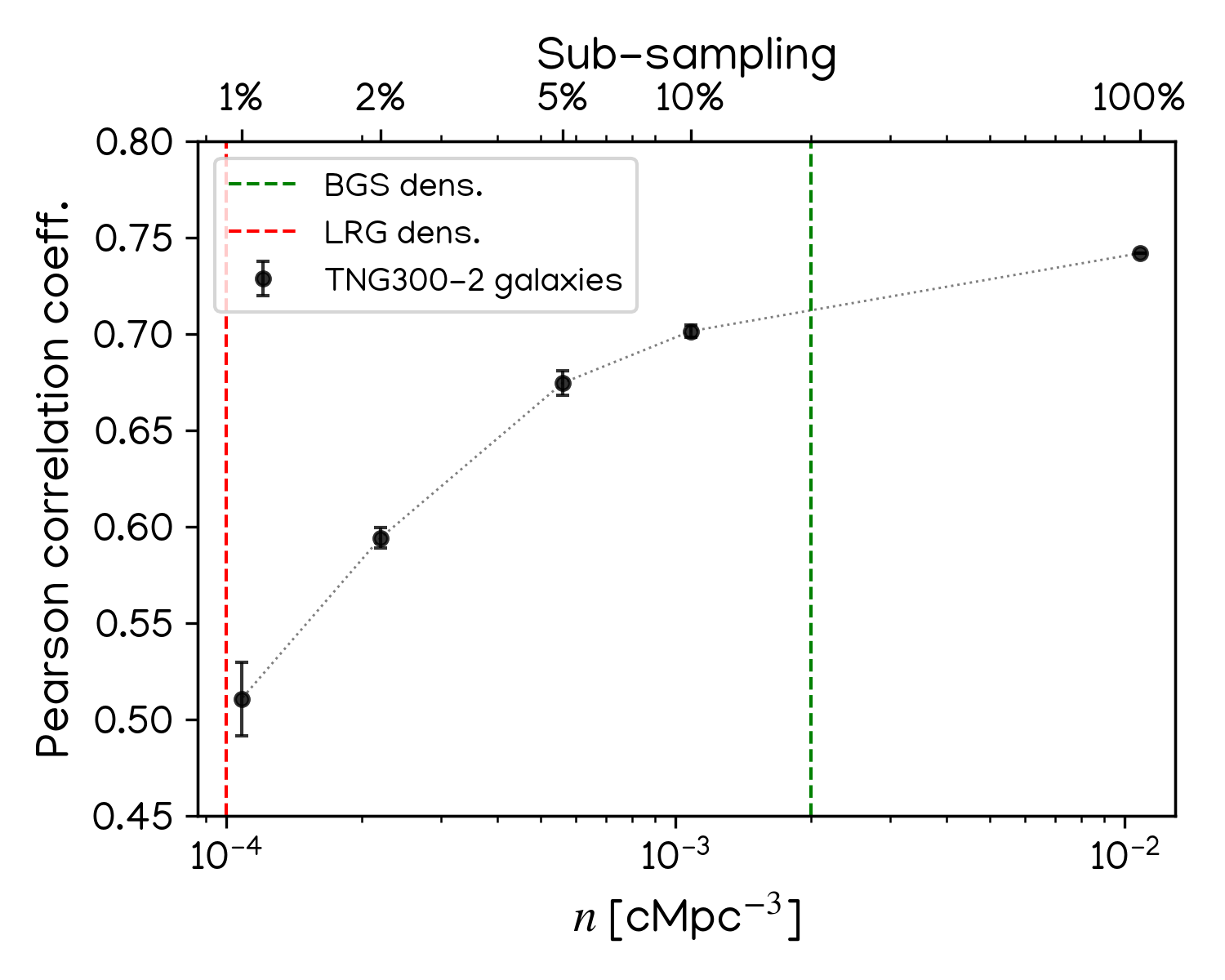}
\caption{Mean Pearson correlation coefficient between real and reconstructed velocities as a function of the number density and sampling fraction of the whole $M_\text{FoF}\in (10^{11}, 10^{12}) \, M_\odot$ \texttt{TNG300$-$2} galaxy sample. Error bars display one standard deviation. Vertical lines display the mean number density of galaxies of the BGS (green) and LRG (red) samples used throughout this work.}
\label{fig:Pearson_coeff}
\end{figure}

\subsection{Convergence test}

To assess the accuracy of the method, we ran our PIZA reconstruction implementation on the galaxies from the \texttt{TNG300$-$2} simulation (see Sect.~\ref{s:bias_in_voids} for a brief description). Galaxy and halo catalogues obtained by the \texttt{Subfind} FoF halo finder \citep{springel2001populating} are publicly available. For the test, we selected the subsample of galaxies with FoF dark matter mass in the $M_\text{FoF}\in (10^{11}, 10^{12}) \, M_\odot$ range, ending up with $N \approx 3\times10^{5}$ matter tracers. Our PIZA reconstruction of velocities was then carried out, taking into account periodic boundary conditions. To calculate the linear bias associated with the data, we adopted the prescription of \citet{tinker2010large}. Assuming the FoF dark matter mass is a good proxy of $M_\text{200m}$ (mass enclosed inside a radius in which the density is 200 times the mean matter density at that time), we ended up with \mbox{$b \approx 0.75$ at $z = 0$} for this halo mass range.

The test results are presented in Fig.~\ref{fig:vel_recon_test}, where we plot contours of the halo reconstructed velocities against real (known) velocities along the three Cartesian directions. A Pearson correlation coefficient of $0.74$ was obtained for the three directions, implying good agreement between the reconstructed and real quantities, given the limitations of the methodology (namely, it cannot model properly beyond shell-crossing). In the three cases, the contours lie on the 1:1 ideal relation, suggesting that our calculation of the linear bias is robust. 

To finally complete the convergence study, we ran the velocity reconstruction algorithm for different sub-samplings of the $M_\text{FoF}\in (10^{11}, 10^{12}) \, M_\odot$ \texttt{TNG300-2} set of galaxies. For each fraction $f_s$ of the whole sample, we produced five different random sub-samplings and calculated the mean Pearson coefficient together with the standard deviation. The test was performed for $f_s = 1, 0.1, 0.05, 0.02$ and $0.01$. The results are displayed in Fig.~\ref{fig:Pearson_coeff}, where the mean Pearson coefficient across all sub-sampling realisations and velocity directions is plotted against the number density of galaxies. While the Pearson correlation coefficient lies between $0.7-0.75$ for number densities in $10^{-3} -10^{-2} \, \mathrm{cMpc^{-3}}$, hence indicating a significant agreement between reconstructed and real velocities, it rapidly decreases for lower values than $10^{-3} \, \mathrm{cMpc^{-3}}$. In particular, for number densities lower than $10^{-4} \, \mathrm{cMpc^{-3}}$ the coefficient goes below 0.5, implying poor correlation from this point.

The BGS sample lies in the number density range in which the PIZA reconstructed velocities display good agreement with real velocities. On the contrary, LRGs lie in the critical zone from which poor correlation starts to be obtained. The accuracy of the dynamical constraints imposed by the void finder when using the velocity divergence directly depends on the accuracy of the underlying velocity field related to the matter distribution used to identify voids. Hence, we expect the BGS void catalogue to be more precise and complete than its LRG counterpart.

\section{Void profiles comparison from observational and mock galaxy samples}
\label{app:obs_mock_equivalence}

\begin{figure}[t!]
\centering 
\includegraphics[width=1\linewidth]{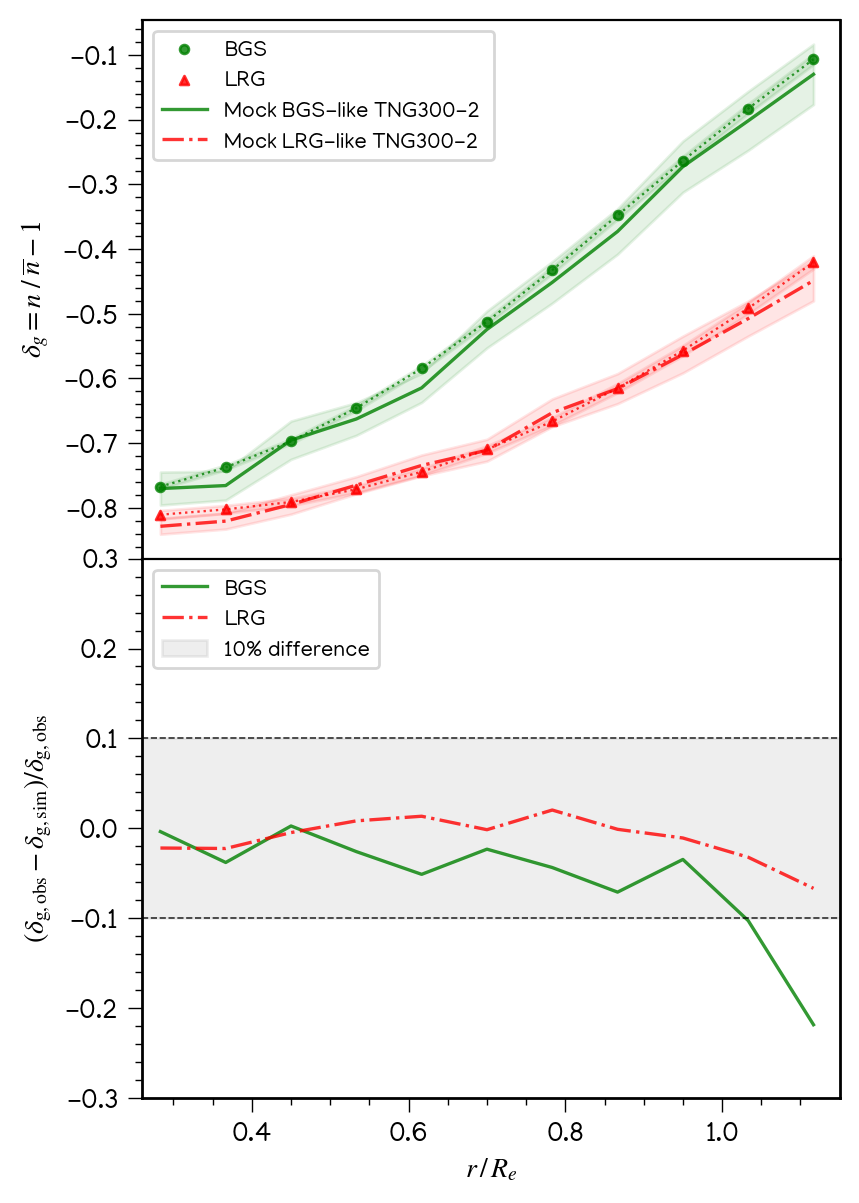}
\caption{Upper panel: Mean galaxy number density profile for both the BGS (green circles) and LRG (red triangles) voids together with their mock BGS-like (solid green line) and LRG-like (dash-dotted red line) profiles obtained from the \texttt{TNG300-2} simulation. The green and red shaded regions represent 1$\sigma$ of the standard error. Lower panel: Relative difference between the observational and mock profiles for both the BGS and LRG cases. The shaded region inside the 10\% relative error is depicted in grey.}
\label{fig:equivalence_profiles}
\end{figure}

To calculate the linear bias inside DESI voids using density profiles \citep[e.g. see][]{pollina2017linearity}, the galaxy number density and the matter density profiles are needed. Since the overall matter distribution in the DESI volume is unknown (note that galaxies are just a biased tracer), such a calculation is, in principle, infeasible. Nevertheless, as described in Sect.~\ref{s:bias_in_voids}, one can follow a simple approach to circumvent this issue: i) first extract two mock galaxy samples mimicking the BGS and LRG redshifts and bias (extracted from the \texttt{TNG300-2} simulation), ii) prove the compatibility between the number density profiles of voids obtained with observational and mock galaxy catalogues, and iii) use the matter density profiles of voids found in the mock galaxy catalogues as an approximation of the real (observational) matter profiles of BGS and LRG voids. 

To justify the compatibility between observational and mock profiles, Fig.~\ref{fig:equivalence_profiles} presents, in the top panel, the mean galaxy number density profiles of BGS (green circles) and LRG (red triangles) voids. In the same panel, we display the same profiles for voids found in the BGS-like and LRG-like mock catalogues with a green solid line and red dash-dotted line, respectively. In the lower panel, the relative difference between both profiles is presented. For $r < R_e$, the relative error lies below 10\% and, hence, in this region, we can reasonably state that the mean matter density profile of the voids found in the mock catalogues (both in the BGS and LRG case) is a proper approximation of the real matter profile that would be obtained if we knew the true matter distribution inside the DESI volume.

\end{appendix}

\end{document}